\documentclass[12pt]{article}
\usepackage{amssymb}
\usepackage{graphicx}
\def\labelmark{}
\def\void{}

{\ifx\void\labelname\def\junk{\end{displaymath}}
\else\def\junk{\end{eqnarray}}\fi\junk\labelmark\def\labelname{}}

\newcommand{\bra}{\begin{array}}
\newcommand{\era}{\end{array}}
\newcommand{\beq}{\begin{equation}}
\newcommand{\eeq}{\end{equation}}
\newcommand{\bqn}{\begin{eqnarray}}
\newcommand{\eqn}{\end{eqnarray}}

\font\mybb=msbm10  at 12pt
\def\bb#1{\hbox{\mybb#1}}

\font\mybbi=msbm10  at 9pt
\def\bbi#1{\hbox{\mybbi#1}}

\def\Z{\bb Z}

\def\BC{\bb C}
\def\_\BC{\bbi C}
\def\RR{\bb R}

\newcommand{\om}{\omega}

\newcommand{\te}{\theta}

\newcommand{\al}{\alpha}

\newcommand{\lga}{\longrightarrow}

\newcommand{\st}{\star}

\newcommand{\ov}{\over}

\newcommand{\no}{\noindent}
\newcommand{\ev}{\equiv}
\newcommand{\lb}{\label}

\begin{document}
\begin{titlepage}
\setcounter{page}{1}
\renewcommand{\thefootnote}{\fnsymbol{footnote}}

\begin{flushright}
UFR-HEP/0203\\
hep-th/0204248
\end{flushright}
%\vspace{3mm}
\begin{center}

{\Large\bf A Matrix Model for
$\normalsize{\nu_{ k_1k_2}={k_1+k_2 \ov k_1 k_2}}$
Fractional \\ Quantum Hall States }

\vspace{5mm}

{\bf Ahmed Jellal$^{1,2,3}$
\footnote{E-mail: {\textsf jellal@gursey.gov.tr }}},
{\bf El Hassan Saidi$^{2}$
\footnote{E-mail: {\textsf H-saidi@fsr.ac.ma}}}
and
{\bf Hendrik B. Geyer$^{3}$
\footnote{E-mail: {\textsf hbg@sun.ac.za }}}
\vspace{3mm}

$^{1}${\em Institut f\"ur Physik,
Technische Universit\"at Chemnitz\\
D-09107 Chemnitz, Germany}\\
%\vspace{2mm}
$^{2}${\em Lab/UFR, High Energy Physics,
Physics Department, Mohammed V University\\
Av. Ibn Battouta, P.O. Box. 1014, Rabat, Morocco}\\
%\vspace{2mm}
$^{3}${\em Institute for Theoretical Physics, University of
Stellenbosch\\
Private Bag X1, Matieland 7602, South Africa}\\

\end{center}

\vspace{3mm}
\begin{abstract}

We propose a  matrix model to describe a class of
fractional quantum Hall (FQH) states for a system
of $(N_1+N_2)$ electrons with filling factor more
general than in the Laughlin case. Our model, which
is developed for FQH states with filling factor of
the form  $\nu_{ k_1k_2}=\frac{k_1+k_2}{k_1k_2}$
($k_1$ and $k_2$ odd integers), has a
$U(N_1)\times U(N_2)$ gauge invariance, assumes that
FQH fluids are composed of coupled branches of the
Laughlin type, and uses ideas borrowed from hierarchy
scenarios. Interactions are carried, amongst others,
by fields in the bi-fundamentals of the gauge group.
They simultaneously play the role of a regulator,
exactly as does the Polychronakos field. We build the
vacuum configurations for FQH states with filling factors
given by the series
$\nu_{ p_1p_2}=\frac{p_2}{p_1p_2-1}$, $p_1$ and $p_2$
integers. Electrons are interpreted as a condensate of
fractional $D0$-branes and the usual degeneracy of the
fundamental state is shown to be lifted by the
non-commutative geometry behaviour of the plane.
The formalism is illustrated for the state at
$\nu=\frac{2}{5}$.

\end{abstract}
\end{titlepage}

\newpage
\tableofcontents
%\newpage

%\vspace{20mm}

%%%%%%%%%%%%%%%%%%%%%%%%%%%%%%%%%%%%%%%%%%%%%%%%%%%%%%%%%
\section{Introduction}
%%%%%%%%%%%%%%%%%%%%%%%%%%%%%%%%%%%%%%%%%%%%%%%%%%%%%%%%%
Susskind's original and suggestive idea that a
non-commutative (NC) U(1) Chern-Simons
theory is the natural effective theory
to approach the fundamental
state of fractional quantum Hall (FQH)
systems~\cite{susskind}, created
an intensely revived interest in
the exploration of new aspects of
FQH fluids~\cite{poly,hellerman1,karabali,
hellerman2,saidi3,kobashi}.
Results on NC
geometry methods and brane systems of $10D$ type-$II$
superstrings~\cite{connes, seiberg} have primarily been used.
Starting from a two-dimensional system with a large number $N$ of
electrons
in the presence of a perpendicular strong
magnetic field $B$, and considering
fluctuations
\beq
\theta \varepsilon^{ij}A_{j}(y)
\eeq
around the time-independent
background $X^{i}=y^{i}$, Susskind showed that
the resulting effective field
theory is a NC $U(1)$ Chern-Simons gauge
theory with $\te={1\ov 2\pi\rho}$
and $\rho$ is the particle density. 

%with an integer level
%\footnote{
%Following~\cite{poly,hellerman1,hellerman2} we know that
%quantum mechanics affects the usual Laughlin state
%filling factor $\nu =\frac{1}{k}$,$\ $
%which should now be read as
%$\nu^{\ast }=\frac{1}{k^{\ast }}$, with $k^{\ast }$
%related to $k$ as $k^{\ast}=k+1$. To implement this
%quantum feature in the Chern-Simons (CS) effective
%field theory model for the FQH ground state,
%the integer level, appearing in
%front of the CS gauge action, should be
%thought of as given by $k^{\ast }$.
%Here, we shall work with the notation
%$\nu =\frac{1}{k}$, but the same
%analysis is a priori valid as well for
%$\nu ^{\ast }=\frac{1}{k^{\ast }}$
%by just substituting $k$ by $k^{\ast }$
%everywhere.}
% In the operator analysis
%for states that are not of Laughlin type,
%a more involved analysis is however needed.}
%\beq
%k=\frac{1}{\nu } \equiv B\theta .
%\eeq
\no The large $U(N)$ automorphism symmetry of the
Susskind matrix model,
\beq
X^{\prime }=UXU^{+}
\eeq
is mapped to an
area-preserving diffeomorphism on the $y$ plane,
$|\frac{\partial^{2}y^{\prime }}{\partial{y^2}}|=1$,
which in turns is mapped to a NC $U(1)$ invariance in
the space of gauge fluctuations,
\beq
A_{i}^{\prime }=U\st (A_{i}-\partial_{i})\st U
\eeq
where $\st$ is the usual Moyal product.

Moreover, by exploring the possibility
to develop a consistent
finite matrix model for the
description of FQH droplet systems, interesting
developments have been made by appropriately treating
the NC finite matrix model constraint
equations. A first development
in this direction was made in~\cite{poly}
where a new field, denoted $\Psi $ and
transforming in the fundamental ${\bf N}$ of $SU(N)$
$({\bar{\Psi}}\sim {\bar{{\bf N}}})$,
has been introduced to regularize the
non-commutative plane constraint equation:
\beq
\lb{const}
[X^{1},X^{2}]=i\theta
\eeq
for the case of a finite number $N$ of electrons.
This equation is consistent only for infinite matrices
$(N=\infty)$, but with the help of the $\Psi $ field,
it can be made consistent even for finite dimensions
as shown below
\beq
\bra{l}
\lb{con}
-iB\lbrack X^{1},X^{2}]_{nm} + \Psi _{n}^{-}\Psi _{m}^{+}
= B\theta \delta _{nm}\\
\sum_{n} \Psi _{n}^{+}\Psi _{n}^{-} = NB\te.
\era
\eeq
With a non-zero $\Psi $ field, the trace
on the states is now well-defined and
the initial constraint equations are turned
to conditions for classical
$U(1)\times SU(N)$ invariance with fixed
$U(1)$ charges as shown in the
second of relations (\ref{con}). There the
$X^{i}$'s are the usual $(0+1)D$ Susskind
matrix field variables transforming as
the adjoint of $U(N)$; their $N$ real
eigenvalues may be thought of as just
the two space coordinates of the $N$
classical electrons. $\Psi $ is an
auxiliary complex scalar field in the
fundamental representation of $SU(N)$,
ensuring consistency of
the finite $N$ restriction.
It can be viewed as the carrier of the boundary
effects in the droplet
approach and turns out to behave as the square
root of the broken $U(1)$
abelian subsymmetry of the original
$U(N)$ invariance of the Susskind matrix
model. Together with the $X^{i}$,
the fields $\Psi $ and ${\bar{\Psi}}$ may
be viewed as following from a unique
$(N+1)\times (N+1)$ matrix field
${Y^{i}}\sim {\rm{\bf Adj}}{(U(N+1))}=
{\rm{\bf Adj}}(U(N))\oplus
{\bf N}\oplus {\bar{{\bf N}}}\oplus {\bf 1}$
where ${\rm{\bf Adj}}(U(N))$ describes the
$X^{i}$ fields while ${\bf N}$ and
${\bar{{\bf N}}}$ describe respectively
$\Psi$ and ${\bar{\Psi}}$. In the
language of $D$-brane physics,
where the particles are viewed as
$D0$-branes dissolved in the $D2$ world volume brane,
the above $\Psi $ field is
represented by a $F1$ string with an end on
a $D0$-brane of $D2$ and the
other end on a
$D4$-brane~\cite{bernevig,brodie,hellerman2},
see also~\cite{kogan,fabinger}.
In section~4, we will explore other aspects
of this field and introduce others in order
to develop models for FQH states
that are not of Laughlin type.

Quantum mechanically, the $X^{i}$ hermitian
matrix variables and the $\Psi $
complex vector are moreover interpreted
as creation and annihilation
(matrix) operators acting on the Hilbert space
${\bf H}$ of states
$\{|\Phi >\}$. In this case,
(\ref{con}) must be understood
as constraint equations that
should be imposed on ${\bf H}$. If one
forgets for a while about the $\Psi $
vector field and focus on the $X^{i}$'s by
setting $Z^{\pm }=(X^{1}\pm iX^{2})$;
then associate with each matrix field
variable $Z_{nm}^{\pm }$, the
$2N^{2}$ harmonic operators
\beq
A_{nm}^{\pm }=\sqrt{B}Z_{nm}^{\pm }
\eeq
obeying the usual Heisenberg algebra,
except that now one has $2N^{2}$
operators
\beq
\bra{l}
\lb{hal}
\left[ A_{nm}^{\mp },A_{kl}^{\pm }\right]
= \pm \delta _{nk}\delta _{ml}
\\
\left[ A_{nm}^{\pm },A_{kl}^{\pm }\right]  = 0.
\era
\eeq
Then the classical constraint equation (\ref{con})
should be replaced by
\beq
\bra{l}
\lb{ncon}
J_{nm}|\Phi >=  0 \\
J_{0}|\Phi >= Nk|\Phi >.
\era
\eeq
In these quantum constraints, the $J_{nm}$
operators, which are expressed in
terms of the $2N^{2}$ harmonic oscillators
$A_{nm}^{\pm }$ and $2N$ harmonic
$\Psi ^{\pm }$ ones as
\beq
J_{nm}=A_{nk}^{-}A_{km}^{+}-A_{nk}^{+}
A_{km}^{-}+\Psi _{n}^{-}\Psi _{m}^{+}
\eeq
define just the usual $SU(N)$ generators while
\beq
J_{0}=\sum_{n} \Psi _{n}^{+}\Psi _{n}^{-}
\eeq
is the charge operator realizing the
abelian $U(1)$ subsymmetry factor of $U(N)$.
As such the quantum constraint
equations (\ref{ncon}) require the wavefunctions
$|\Phi >\ \in {\bf H}$ \ to be $SU(N)$
invariant and moreover carry $Nk$ charges
of $U(1)$; that are having $Nk$
operators of type $\Psi ^{+}$ or, equivalently,
a monomial form $\left( {\Psi ^{+}}\right)^{kN}$.
In~\cite{hellerman1}, see also~\cite{hellerman2},
the solution for the constraint
equations (\ref{con}) have been obtained by using special
properties of antisymmetric and
holomorphic polynomials and the vacuum
configuration has been shown to be
similar to that obtained years ago by
Laughlin~\cite{laughlin}.

Despite the success of the Susskind
NC model and its regularised version introduced
by Polychronakos, in particular
the theoretical prediction $\te={1\ov 2\pi\rho}$
and the recovering of the
Laughlin wavefunctions,
several open questions remain
which are not addressed by the Susskind approach.
One of these questions concerns
FQH states that are not
of the Laughlin type.
%, that is FQH states at filling factor $\nu \neq \frac{1}{k}$. 
In fact there are many FQH states,
such as $\nu =\frac{2}{3},\frac{2}{5}, \frac{3}{7},\dots $,
that have been observed experimentally~\cite{prange}
but are not recovered by the Susskind model.

Another question, which has not been
addressed even for the case of the
Laughlin fluid, concerns the singularities
of the Laughlin wavefunctions
$\Phi _{L}(z_{1},\dots ,z_{N})=<z_{1},
\dots ,z_{N}|\Phi _{L}\rangle $ with
filling factor $\nu =\frac{1}{k}$.
These wavefunctions
\beq
\lb{lau}
\Phi _{L}(z_{1},\dots,z_{N})=\prod_{\al<\beta =1}^{N}
\left( z_{\al}-z_{\beta}\right)^{k}\exp \left(-\frac{B}{4}
\sum_{\sigma }\left| z_{\sigma }\right| ^{2}\right)
\eeq
have a huge ${\Z}_{k{N(N-1)\ov 2}}$ discrete symmetry
containing the special ${\Z}_{k}^{N(N-1)\ov 2}
$ and ${\Z}_{N(N-1)\ov 2}^{k}$ subsymmetries
and degenerate zeros of degree $k$.
These zeros are expected to play a crucial
role in the study of the quantum
configuration of the NC system. Recall
in passing that the degenerate zeros
of $\Phi _{L}$ cover remarkable features
which may be exploited in the
analysis of the quantum spectrum of the
FQH system with fractional values for
the filling factor. Indeed, setting
\beq
\lb{set}
\bra{l}
u=\Phi _{L}(z_{1},\dots ,z_{N})\\
v_{\al\beta}=\prod_{\gamma <\delta \neq
\left( \alpha ,\beta \right) }^{N}\left(
z_{\gamma }-z_{\delta }\right)^{-k}\exp
\left(\frac{B}{4}\sum_{\sigma }\left|
z_{\sigma }\right| ^{2}\right) \;
\era
\eeq
so their product can be written as
\beq
\lb{nlau}
uv_{\al\beta}=\left(z_{\alpha }-z_{\beta }\right)^{k}
\eeq
which is nothing but the usual
$SU\left( k\right) $ singularity equation
of the assymptotically locally Euclidean (ALE)
space~\cite{berenstein1,belhaj}.
As such one expects that many results obtained in
the context of representation
theory for NC manifolds with
singularities~\cite{berenstein2,saidi}
can be applied as well to the FQH
systems. From the NC geometry point
of view, the above mentioned degeneracy
of $\Phi _{L}(z_{1},\dots ,z_{N})$
should be lifted and one expects to
get richer solutions for vacuum configurations
in the NC plane that should
contain the one recently built
in~\cite{hellerman1}.

The aim of this paper is to develop a matrix
model for the FQH states at
filling factor given by the series
\beq
\lb{jsf}
\nu_{k_1k_2} =\frac{k_{1}+k_{2}}{k_{1}k_{2}}
\eeq
where $k_{1}$\ and $k_{2}$\ odd
integers and work out the vacuum
configurations by taking into account
the singularities of the
Laughlin wavefunctions and the NC
geometry of the plane.
To fix the ideas, we will mainly
focus our attention on the FQH states at
filling factor $\nu =\frac{2}{5}$. First,
we reconsider the Laughlin states with
$\nu =\frac{1}{k}$ and study
the effective link between discrete symmetries
and the NC geometry of the
plane. We then look for the general
solutions for vacuum wavefunctions
by using techniques, borrowed from
non-perturbative QCD concerning
compositeness~\cite{becher}.
It is then possible to determine vacuum
configurations for the wavefunctions as
suggested by NC geometry of the plane.
This also allows us to propose a
way of thinking about the electrons of FQH
states as condensate states,
formally similar to baryons of hadronic
models of strong interactions at low
energies, and to the elementary excitations
as the fundamental constituents
analogously to quarks in QCD.
In brane language electrons are
represented by $D0$-branes, while
the elementary excitations appear as fractional
$D0$-branes. This quantum description recovers
not only the Susskind
construction, but also the Hellerman
and Van Raamsdonk solution for the
constraint equations (\ref{con}),
and of course the Laughlin wavefunctions with zeros of
order $k$.

We then consider quantum configurations
for states that are of
non-Laughlin type by using, on one hand,
the developments made in the
framework of Susskind's proposal and
the subsequent results and, on the other
hand, taking advantage from a special feature
of the continuous fraction
to interpret FQH states as a system
of coupled Laughlin states.
Recall that ideas utilizing coupled
Laughlin states to describe states
with general filling factors were successfully
implemented in the past when
studying abelian hierarchies and effective
Chern-Simons (CS) gauge models~\cite{wen}.
Such analyses were based on the action
\begin{equation}
\lb{csg}
S[A^{1},\dots ,A^{l}]=\frac{1}{4\pi }
\int d^{3}y\; \varepsilon^{\mu
\nu \rho }\ K_{IJ}
\partial _{\mu }A_{\nu }^{I}A_{\rho }^{J}.
\end{equation}
In the present description the world volume
of the $D2$-brane, where the $A^{I}$
CS gauge fields propagate, have a fibration
${\bf B}\times {\bf F}$ of a base
given by the world volume of the $D2$-brane
and as a fiber the vector space
${\bf E}$ generated by the $\{{\bf r}_{I}\}$
vector basis system such that
${\bf r}_{I}\ {\bf r}_{J}=K_{IJ}$.
In this basis the hierarchical gauge field
components $A^{I}$ appear just as projections
on the vector basis, $A_{j}^{I}={\bf r}_{I}\ {\bf A}_{j}$.
The $K_{IJ}$ matrix appearing in the
above action functional is the well-known
matrix topological order
describing hierarchies; for details on
the possible classes of $K_{IJ}$,
see~\cite{wen}.

The organization of this paper is as follows.
In section 2, we develop a
microscopic analysis of the Hamiltonian
description of the Laughlin states
with filling factor $\nu =\frac{1}{k}$.
We study the resolution of the
$SU\left( k\right) $ singularity by NC
geometry and work out the resulting
wavefunctions describing the vacuum
configuration. In section 3, we study
the FQH states that are not of Laughlin
type and propose a way to approach
such states by using the earlier Susskind results
and ideas on fluid branches. In
section~4, we consider a system of
$\left( N_{1}+N_{2}\right)$ electrons
and develop a matrix model for FQH states
with filling factor given by the
series $\nu_{k_1k_2}$. This system contains two
branches; a basic one with $\nu _{k_1}=\frac{1}{k_{1}}$
and another one with $\nu _{k_2}=\frac{1}{k_{2}}$ built
on the top of the $\nu _{k_1}$
state. The coupling of the two branches is
ensured by the introduction of an
effective $B^{\ast }$ field and
moreover through the use of a bosonic
field in the bi-fundamental of the
$SU\left( N_{1}\right) \times SU\left(N_{2}\right) $
group. Such a field turns out not only to carry
the interaction, but also to play the role of
a regulator without even requiring Polychronakos
type fields.

%%%%%%%%%%%%%%%%%%%%%%%%%%%%%%%%%%%%%%%%%%%%%%%%%%%%%%%%%
\section{Microscopic description}
%%%%%%%%%%%%%%%%%%%%%%%%%%%%%%%%%%%%%%%%%%%%%%%%%%%%%%%%%

To start recall that the Lagrangian $L$
describing the dynamics of an
electron of mass $m$ and charge $(e=1)$
moving in two dimensional
space ${\bf x}=(x^{1},x^{2})$ in the presence
of a perpendicular external
constant magnetic field $B$ and a potential $U(x)$,
($U(x)={\frac{\kappa }{2}}{\bf x^{2}}$,
$\ $with $\kappa $ is a coupling constant) is
\beq
\lb{lag}
L={\frac{{m}}{2}}\dot{{\bf x}}^{2}+
{\frac{B}{2}}\dot{{\bf x}}\wedge
{\bf x}-{\frac{\kappa }{2}}{\bf x}^{2}.
\eeq
The Hamiltonian
\beq
H={\bf \pi }\dot{{\bf x}}-L
\eeq
of this electron is obtained as usual by
computing the conjugate
momenta ${\bf \pi }=\frac{\partial L}
{\partial \dot{{\bf x}}}$.
% of the $ z^{\pm }=(x^{1}\pm ix^{2})$ positions.
In the presence of a strong enough
magnetic field $B$, 
%these momenta are
%proportional to the positions and reduce
%to
%\beq
%\bra{l}
%\pi =i\frac{B}{2}z\\
%{\bar{\pi}}=-i\frac{B}{2}{\bar{z}}.
%\era
%\eeq
%The Hamiltonian $H\left( z^{+},{z}^{-}\right)
%\equiv H\left( z,{\bar{z}}
%\right) $ thus emerges as the
%one dimensional harmonic oscillator
%\beq
%\lb{hamo}
%H={\frac{\kappa }{2}}z{\bar{z}}.
%\eeq
%One of the remarkable features of~(\ref{hamo}),
%which can equivalently be written as
%$\mathcal{H}={\frac{\kappa }{4}}
%\left( \overline{z}{z+z}\overline{z}\right)$,
%is that it depends only on the complex space
%coordinates $z$ and $
%\overline{z}$, but no time derivatives of variables.
%This property has a very attractive
%interpretation at the quantum level:
%the real plane is quantized into
%small cells of area proportional to the
%inverse of the magnetic field $B$.
%Indeed, extending the usual construction
%for the case of a quantum
%oscillator in the $\left( x,p\right) $ phase space,
the quantum Hamiltonian $\mathcal{H}$ may be defined as
\beq
\lb{qham}
\mathcal{H}={\frac{\kappa }{4}}\left( \bar{Z}Z{+Z}\bar{Z}\right)
\eeq
where now $Z\equiv Z^{-}$ and 
${\bar{Z}\equiv Z}^{+}$ are the Heisenberg
operators associated to $z$ and 
${\bar{z}}$. This one particle energy
operator may be viewed as describing 
the energy configurations of a harmonic
oscillator with $a^{\pm }$ creation 
and annihilation operators:
\beq
\bra{l}
a^{+}=\sqrt{B}{\bar{Z}}\\
a=\sqrt{B} Z
\era
\eeq
satisfying the usual commutation relations, namely
\beq
\bra{l}
\left[ a^{-},a^{+}\right] =1\qquad \\
\left[ a^{\pm },a^{\pm }\right] =0.
\era
\eeq
In terms of these operators, (\ref{qham}) reads as
\beq
\mathcal{H}={\omega\ov 2} \Big(2a^{+}a+1\Big)
\eeq
where $\omega ={\frac{\kappa }{B}}$ and
with $\hbar =1=c$. For $\kappa $ large enough, say
$\kappa =\frac{B^{2}}{4}+\kappa ^{\prime }$,
where $\kappa^{\prime }$\ is some perturbation
parameter, the gap energy is large
($E_{1}\sim \kappa \sim B$); and the dynamics
of the quanta is mainly given by
oscillations near the origin within
the lowest Landau level (LLL).

Though standard, the analysis we presented 
above yields some valuable
information about the discrete nature of 
the real plane, induced by
the $B$
field at the quantum level. Perhaps 
the most important
piece of information one obtains 
comes from the relation
\beq
\left[ Z^{-},Z^{+}\right] =\frac{1}{B}
\eeq
which tells us that, from the semi-classic
point of view, everything appears as if the 
$(z,{\bar{z}})$ plane, $z=x_1+ix_2$, 
is quantized in fundamental
$B$ dependent areas (say
small squares or discs),
\beq
A_{0}=\frac{2\pi }{B}
\equiv l_{0}^{2}
\eeq
where the magnetic length $l_{0}$ appears 
in our notational
convention as just the fundamental 
length of the edges of the small
square.
\subsection{$\nu =\frac{1}{k}$ FQH states and discrete symmetries}
%%%%%%%%%%%%%%%%%%%%%%%%%%%%%%%%%%%%%%%%%%%%%%%%%%%%%%%%%%%%%%%%%%
From the semi--classic point of view,
and due to the presence of the
magnetic field $B$, the space coordinate
$z_{\alpha }$, parameterizing an
electron in the plane, should be thought of
as a $k\times k$ matrix of the $k$--dimensional 
representation $\mathcal{D}({\Z}_{k})$
of the group ${\Z}_{k}$. The full NC plane
is indeed a kind of fibration $\mathbf{B}\times
\mathbf{F}$ whose base $\mathbf{B}$ is
a plane with $\mathbf{F}$ as a fiber
$\mathcal{D}({\Z}_{k})$.
This property is a well-known
feature in constructing a NC geometry
extension of manifolds with $SU(k)$
singularities~\cite{berenstein1,belhaj,berenstein2,saidi}.
A quasi-similar situation happens
here in the study of the vacuum
configurations of FQH fluids from a
non-commutative point of view. We will
first describe the discrete
symmetries of the Laughlin wavefunctions and
then develop the basis of a NC analysis for FQH fluids.

%%%%%%%%%%%%%%%%%%%%%%%%%%%%%%%%%%%%%%%%%%%%%%%%%%%%%
 {\bf{Symmetries}:}
%%%%%%%%%%%%%%%%%%%%%%%%%%%%%%%%%%%%%%%%%%%%%%%%%%%%%

\no To exhibit the discrete symmetries
of (\ref{lau}) more clearly, recall
first that the Laughlin wavefunctions
of filling factor $\nu =\frac{1}{k}$
is completely antisymmetric under the
changes of any pair of electrons of
coordinates $z_{\alpha }$ and $z_{\beta }$,
provided $k$ is an odd integer, that is
\beq
\Psi _{L}\left( z_{1,...,}
z_{\alpha },...,z_{\beta },...z_{N}\right) =\left(
-\right) ^{k}\Psi _{L}\left( z_{1,...,}
z_{\beta },...,z_{\alpha},...z_{N}\right).
\eeq
These wavefunctions also have a
${\Z}_{k\frac{N\left( N-1\right) }{2}}$
manifest discrete invariance containing
${\Z}_{\frac{N\left(N-1\right) }{2}}^{k}$
and ${\Z}_{k}^{\frac{N\left( N-1\right) }{2}}$
as two special subsymmetries with a
remarkable interpretation. These
symmetries are directly seen in
$\Psi _{L}\left( z_{1}...z_{N}\right)$ by
requiring invariance under the
change of variables
\beq
\bra{l}
z_{\alpha }^{\prime } =\lambda z_{\alpha }\\
\Psi _{L}^{\prime } = \Psi _{L}
\left( z_{1}^{\prime }...z_{N}^{\prime}\right)\; 
\era
\eeq
where $\lambda $ is a complex (group) parameter
and $\alpha =1,...,N$.
The condition we get from the identity
$\Psi _{L}^{\prime }=\Psi _{L}$ is the
constraint equation on $\lambda $:
\beq
\lb{lam}
\lambda ^{k\frac{N\left( N-1\right)}{2}}=1.
\eeq
This is a constraint relation which,
in general, has several solutions
describing different subgroups of the
huge periodic symmetry
${\Z}_{k\frac{N\left( N-1\right) }{2}}$,
parameterized by the fundamental
root parameter
\beq
\lb{lambda}
\lambda _{0}=\exp \Big[i\frac{4\pi }
{kN\left( N-1\right) }\Big].
\eeq
The two particular solutions of~(\ref{lam})
we here refer to above are those
associated with the two special subsymmetries
${\Z}_{\frac{N\left(N-1\right) }{2}}$ and
${\Z}_{k}$, respectively, generated by
\beq
\xi=\lambda _{0}^{k}=\exp i\frac{4\pi }
{N\left( N-1\right) }
\eeq
and
\beq
\eta=\lambda _{0}^{\frac{N\left( N-1\right) }{2}}=
\exp (i\frac{2\pi }{k}).
\eeq
Before proceeding we want to make two comments
regarding these invariances.
(i) First, the integer $\frac{kN\left( N-1\right) }{2}$
appearing in ~(\ref{lam})
is just the total angular momentum of
the Laughlin states; it is a positive
integer multiple of $k$. From the expression
of the Laughlin wavefunctions,
which for $k=1$ reduces essentially to
the Slater determinant
${\rm det}_{\rm Slater}\equiv \Psi _{L}^{\left( k=1\right) }$,
one recognizes the ${\Z}_{\frac{N\left( N-1\right) }{2}}$
invariance as just the symmetry of the
integer quantum Hall (IQH) state. 
%From this point of view,
%${\Z}_{k\frac{N\left( N-1\right) }{2}}$
%is an enhanced symmetry of
%${\Z}_{\frac{N\left(N-1\right) }{2}}$,
%required to implement fractionality of the filling
%factor of the Laughlin wavefunctions when
%starting from an integer one.
(ii) Concerning the ${\Z}_{k}$
subsymmetry of the Laughlin
ground states, it is interesting
to note that it has no analogue
for IQH states, as
the latter are essentially
described by the Slater determinant
and reflects the degeneracy
property of the zeros of $\Psi _{L}$.
This subsymmetry is then
ineherent to the fractional feature
of the filling factor of
ground states of the Hall system; it
is expected to encode valuable
information on Laughlin states with
$\nu =\frac{1}{k}$. It is this aspect
that we now continue to explore.

%%%%%%%%%%%%%%%%%%%%%%%%%%%%%%%%%%%%%%%%%%%%%%%%%%%%%
   {\bf{Degenerate zeros}}:
%%%%%%%%%%%%%%%%%%%%%%%%%%%%%%%%%%%%%%%%%%%%%%%%%%%%%

\no A glance at the structure of the Laughlin wavefunctions lets one
discover that fractionality of the electric charge and the spin
of the quasiparticles which one encounters in the framework of the
Chern-Simons effective field model, is associated with the
degenerate zeros of $\Psi _{L}$; i.e.
\beq
\Psi _{L}\sim \prod_{\alpha
<\beta }^{N}\left( z_{\alpha }-z_{\beta }\right)^{k}.
\eeq
This behaviour of $\Psi_{L}$, which is very
familiar in the physics of quantum systems
living on orbifolds, resembles a similar
situation that one has in singularity theory,
especially the $SU\left(k\right)$ singularity
of the ALE space. To explore the idea, let us,
for simplicity of the analysis, set for a
moment $N=2$ so that $\Psi_{L}\equiv \Psi$
reduces to the monomial
\beq
\lb{tplau}
\Psi \sim \left(z_{1}-z_{2}\right)^{k}.
\eeq
Under the special ${\Z}_{k}$
symmetry, with $\eta =\exp (i\frac{2\pi }{k})$
and $\eta^{k}=1$,
as described above, we have
\beq
\bra{l}
\left( z_{1}^{\prime }-z_{2}^{\prime }\right) =\eta \left(
z_{1}-z_{2}\right) \\
\Psi ^{\prime } =\Psi.
\era
\eeq
Now, expressing $\Psi$ as
\beq
\Psi =\frac{u}{y}\; 
\eeq
with the following
properties under ${\Z}_{k}$ symmetry
\beq
\bra{l}
u^{\prime }=\lambda u\\
y^{\prime }=\lambda y\\
\frac{u^{\prime }}{y^{\prime }}=\frac{u}{y}\; 
\era
\eeq
and substituting in~(\ref{tplau}), we get, by renaming $v=y^{-1}$,
the well-known equation of the ALE space with an ordinary
$SU\left( k\right) $ singularity, namely
\beq
uv=\left( z_{1}-z_{2}\right) ^{k}.
\eeq
Of course this relation can be obtained
from~(\ref{nlau}) by fixing $y=1$ and $N=2$. Therefore 
for generic values of $N$, it extends
straightforwardly and is basically as in~(\ref{nlau}).
${\Z}_{k}$ invariance reflects then the fact
that the Laughlin wavefunctions have zeros with order of
degeneracy $k$. This is the algebraic way
in which the information about the
fractionality of the filling factor $\nu =\frac{1}{k}$ is encoded.
Since we are interested precisely in this behaviour of the quantum
Hall system, let us extend
this treatment by looking for solutions of the
transformations
\beq
z_{\alpha }^{\prime }=\mathbf{P}_{\eta }\
z_{\alpha }\ \mathbf{P}_{\eta }^{-1}
\eeq
and
\beq
\mathbf{P}_{\eta }\Psi_{L}=\Psi _{L}
\mathbf{P}_{\eta }\; 
\eeq
or simply
\beq
z_{\alpha}^{\prime }=\eta_{\alpha }z_{\alpha }\; 
\eeq
which can also be written as
\beq
\lb{trans}
\mathbf{P}\ z_{\alpha }=\eta _{\alpha }\
 z_{\alpha }\ \mathbf{P}\; 
\eeq
where we have denoted the ${\Z}_{k}$
generator $\mathbf{P}_{\eta }$ simply as $\mathbf{P}$.
Our idea of considering the above relation as an additional
constraint equation is borrowed from the
analysis developed in the context of a
resolution of stringy singularities through
non-commutative geometry and discrete
torsion. In this approach, originally
due to Berenstein and Leigh, $\mathbf{P}$
is no longer seen as the generator of an external automorphism
${\Z}_{k}$ symmetry, as we have been doing until now, but as
generating inner automorphisms. For more
details on this issue see~\cite{belhaj,berenstein2,saidi}.
For present purposes we only need to note
that the $\ z_{\alpha }$ variables of~(\ref{trans})
may be solved using fiber bundle techniques
with the plane as base and the $k$-dimensional
representation of ${\Z}_{k}$ as the fiber.
The trivial realization of $z_{\alpha }$ is
given by the tensor product
$z_{\alpha }=w_{\alpha }\otimes \mathbf{Q}$,
where the $w_{\alpha }$'s are
the effective complex plane coordinates
of the electrons and where $\mathbf{Q}$ is
the generator of the internal structure which, together
with $\mathbf{P}$, are realized as
\beq
\bra{l}
{\bf P} =\sum_{j=1}^{k}\
\eta ^{j-1}{\bf \Pi }_{j}\\
{\bf P}^{k}={\bf I}_{k} \\
{\bf Q} =\sum_{j=1}^{k}{\bf f}_{j}\\
{\bf Q}^{k}={\bf I}_{k}.
\era
\eeq
In these relations
\beq
{\bf \Pi }_{j}=|j><j|
\eeq
is the projector on the
$j$-th state of the vector basis
${\cal B}\equiv \{|j>;\ 1\leq j\leq k\}$
of the ${{\cal D}({\Z}_{k})}$ representation,
\beq
{{\bf f}_{j}}=|j+1><j|
\eeq
is a translation operator and ${\bf I}_{k}$
is the usual identity matrix.
The $k\times k$ matrix ${\bf Q}$ is then
a translation operator rotating the
elements of the ${\cal B}$ basis and acting
as ${{\bf f}_{j}}$. The matrix $
{\bf P}$, which is diagonal, may also be written as
\beq
{\bf P}=\exp(i\frac{2\pi L_{0}}{k})
\eeq
where $L_{0}$
is the hermitian operator counting the
${\Z}_{k}$ charges, that is, acting as
\beq
L_{0}|j\rangle =j|j\rangle
\eeq
so that
\beq
{\bf P}|j\rangle =\eta ^{j}|j\rangle.
\eeq
Straightforward calculations show
moreover that ${\bf P}$ and ${\bf Q}$ satisfy
\begin{equation}
{\bf P}{\bf Q}=\eta \ {\bf Q}{\bf P}.
\end{equation}
At the quantum level, the solution
$z_{\alpha }^{\pm }=w_{\alpha }^{\pm}\otimes
{\bf Q_{\alpha }^{\pm }}$ for the non-commutative
point should be naturally replaced by
\beq
Z_{\alpha }^{\pm }=W_{\alpha }^{\pm }\otimes
{\bf Q}_{\alpha }^{\pm }.
\eeq
For the special case
$k=1$, there is no singularity and
the $Z_{\alpha }^{\pm }$ position operators
reduce to the complex plane
operators; i.e. $Z_{\alpha }^{\pm }=W_{\alpha }^{\pm }$,
and so one is left
with a system of IQH states at
filling factor $\nu =1$. In this case the
creation and annihilation operators
$a_{\alpha }^{\pm }=\sqrt{B}$ $Z_{\alpha}^{\pm }$
may be thought of as the operators associated with IQH states. For
later use we shall refer to the
$a_{\alpha }^{\pm }$ with $k=1$ as
$c_{\alpha }^{\pm }$; i.e.
$\left( a_{\alpha }^{\pm }\right) _{k=1}=c_{\alpha}^{\pm }$
and keep the notation $a_{\alpha }^{\pm }$ for creation and
annihilation operators in the NC plane. For the generic case
$k\geq 2$, the filling factor is no longer integer and
the creation and annihilation
operators carry an internal structure induced
by the ${\Z}_{k}$ symmetry
as shown on $a_{\alpha i_{\alpha }}^{\pm }$.
The extra index $i_{\alpha }$
refers effectively to this internal feature and
its realization may be made
more explicit in the trivial
representation with the aid of the ${{\bf f}_{j}}
$ step operators on the internal space, introduced earlier as
\begin{equation}
\lb{real}
a_{\alpha i_{\alpha }}^{\pm }=
c_{\alpha }^{\pm }\otimes {{\bf f}_{i_{\alpha}}}.
\end{equation}
Taking the sum over all states of the internal
space, namely
\beq
\sum_{i_{\al}=1}^{k}a_{\alpha i_{\alpha }}^{\pm }=
c_{\alpha }^{\pm }\otimes {\bf Q}^{\pm }\; 
\eeq
which we set as $a_{\alpha }^{\pm }$ for simplicity,
one sees that the above
operators exhibit a set of special
features, the main ones being: (i) (\ref{real})
reflects a well-known property in brane physics
in the presence of a $B$ field,
namely the fractionating of $D$-branes
at singularities. As such creation
and annihilation operators $a_{\alpha }^{\pm }$
of the electron ( $D0$-brane) fractionate
in terms of more fundamental operators
$a_{\alpha i_{\alpha }}^{\pm }
=\rm{Tr}({{\bf f}_{i_{\alpha }+1}}a_{\alpha }^{\pm })$
(fractional $D0$-branes). This means that in
the same manner that the electrons are described
by $D0$-branes in the brane picture, the
$a_{\alpha i_{\alpha }}^{\pm }$ are associated
with fractional $D0$-branes
(fractional electrons or, again, quasi-electrons).
(ii) The $a_{\alpha }^{\pm }$ operators
carry a ${\Z}_{k}$ charge\ equal to
$\left( \pm 1\right) $ as shown
below:
\beq
\bra{l}
{\bf P}\ a_{\alpha }^{\pm }
{\bf P}^{-1} =\eta ^{\pm }a_{\alpha }^{\pm }\; 
\\
{\bf P}\ A_{\alpha }^{\pm }
{\bf P}^{-1} =A_{\alpha }^{\pm }
\era
\eeq
while the ${\Z}_{k}$ \ scalars are mainly
given by composite operators of the
form
\beq
A_{\alpha }^{\pm }\sim {(a_{\alpha }^{\pm })^{k}}=
{(c_{\alpha }^{\pm})^{k}}\otimes {\bf I}_{k}
\equiv C_{\alpha }^{\pm }\otimes {\bf I}_{k}.
\eeq
To get the right expressions of the invariant
$A_{\alpha }^{\pm }$ bounds in
terms of the $a_{\alpha i_{\alpha }}^{\pm }$'s,
recall that the commutation relations describing
the quantum behaviour of the FQH states at filling
factor $\nu =\frac{1}{k}$ read as
\beq
\bra{l}
\left[ {\bf a}_{\alpha i_{\alpha }}^{-},
{\bf a}_{\beta j_{\beta }}^{+}\right]
=\delta _{\alpha \beta } \delta _{i_{\alpha}j_{\beta}} \\
\left[ {\bf a}_{\alpha i_{\alpha }}^{\pm },
{\bf a}_{\beta j_{\beta }}^{\pm }
\right] =0.
\era
\eeq
The total Hamiltonian
\beq
{\cal H}=\sum_{\alpha =1}^{N}{\cal H}_{\alpha}
\eeq
of the system is given by the following
$k\times k$ diagonal matrix operator
\begin{equation}
\lb{tham}
{\cal H}={\omega\ov 2} \left(\sum_{\alpha =1}^{N}
\sum_{i_{\alpha }=1}^{k} 2 {\bf a}_{\alpha i_{\alpha }}^{+}
{\bf a}_{\alpha i_{\alpha }}^{-}+ N \right).
\end{equation}
${\cal H}$ is proportional to the
${\bf I}_{k}$ identity and has a
manifest ${\Z}_{k}\subset SU(k)$ invariance.
Since
\beq
\left[ {\bf L}_{0},
{\bf a}_{\alpha i_{\alpha }}^{\pm }\right] =
\pm {\bf a}_{\alpha i_{\alpha }}^{\pm }\; 
\eeq
the ${\bf L}_{0}$
charge commutes with ${\cal H}$
and so the vacuum state $|v>$ is degenerate;
it is an eigenstate of both $
{\cal H}$ and ${\bf L}_{0}$;\ that is
\beq
{\cal H}|v>=E_{0}|v>
\eeq
and
\beq
{\bf L}_{0}|v>=j|v>\; 
\eeq
where $1\leq j\leq k$.
In other words, the vacuum $|v>$ is a
vector $|0,j>$ of the $k$-dimensional
representation ${\cal D}\left({\Z}_{k}\right)$
of the group ${\Z}_{k}$. However as
${\cal D}\left({\Z}_{k}\right)$ is
completely reducible
into one-dimensional spaces, one can choose
only one vacuum, say the ${\Z}_{k}$
invariant one, namely $|0>\equiv |0,1>$,
and carry out the usual procedure to
build excited states of~(\ref{tham}), but
keeping in mind that the same analysis
may be done for the $\left(k-1\right)$
others. The rotation between the different
spectra is ensured
by the outer-automorphism ${\bf Q}$
\beq
\bra{l}
{\bf Q}|0,j >=|0,j+1>\ \ \rm{\it modulo}\; k \\
{\bf Q}{\cal H} ={\cal H}{\bf Q}.
\era
\eeq

Another remarkable property concerning the
$\Z_{k}$ symmetry follows from the
obvious identity
\beq
{\bf f}_{i}|j>=|i+1>\ \ {\it if}\; i=j
\eeq
and zero otherwise. This
implies in turn that
\begin{equation}
\lb{rema}
{\bf a}_{\alpha i_{\alpha }}^{\pm }
|n_{\alpha }, j_{\alpha}>\; \sim \;
\delta_{i_{\alpha}j_{\al}}|n_{\alpha }\pm 1, i_{\alpha }+1>.
\end{equation}
An equivalent statement is that for
$k\geq 2$, and due to the property
${\bf f}_{i}^{2}=0$ and ${\bf f}_{i+s}{\bf f}_{i}=0$
for all values of $s\neq 1$
{\it modulo} $k$, the creation and annihilation
operators ${\bf a}_{\alpha i_{\alpha }}^{\pm }$
fulfil very special features, mainly inherited from
those of the ${\bf f}_{i}$'s as can be
seen from~(\ref{rema}):
\beq
\bra{l}
\lb{prop}
\left( {\bf a}_{\alpha i_{\alpha }}^{\pm }\right) ^{2} =0
\\
{\bf a}_{\alpha, \left( i_{\alpha }+s\right) }^{\pm }\cdot
{\bf a}_{\alpha i_{\alpha }}^{\pm } =0\ \ \rm{\it
unless}\,\; s = 1.
\era
\eeq
Due to these identities, one can show
that one can build out of the
${\bf a}_{\alpha i_{\alpha }}^{\pm }$'s a few
${\Z}_{k}$\ invariant composite
operators; $k$ condensates
$(A_{\alpha }^{+})_{(j)}$, $1\leq j\leq k$, given
by the following ordered product with a non-zero
action on the $|j>$ state only
\beq
\bra{l}
\lb{stat}
\Big(A_{\alpha }^{+}\Big)_{(j)} =a_{\alpha ,k+j-1}^{+}
a_{\alpha ,k+j-2}^{+} \dots a_{\alpha ,j+1}^{+}
a_{\alpha ,j}^{+} \\
{\bf P}\cdot \Big(A_{\alpha }^{+}\Big)_{(j)} =
\Big(A_{\alpha }^{+}\Big)_{(j)}\cdot {\bf P}.
\era
\eeq
Under ${\bf Q}$ action these
operators $(A_{\alpha }^{+})_{(j)}$ are
rotated among themselves and under ${\bf Q}^{m}$
they are mapped to $(A_{\alpha }^{+})_{(j+m)}$.
Another ${\Z}_{k}$ invariant composite operator is
given by the trace $\rm{Tr}(A_{\alpha }^{+})_{(j)}$;
this operator does not
depend on the $\left\{ |j>\right\} $
basis vectors and as we show
later, this is the operator that has been used
in~\cite{hellerman1} to construct the
wavefunction. To build the generic
eigenstates
$|\Phi \rangle=|\{n_{\alpha ,i_{\alpha }}\}\rangle $
of the Hamiltonian ${\cal H}$, one
proceeds as usual by acting by monomials
in the creation operators on the
vacuum. This is a standard analysis which we
will skip and come directly to
the study of the case of a FQH system of
$N$ particles with filling factor
$\nu =\frac{1}{k}$. To get the fundamental
state of this system of $N$ particles,
one should solve
the following constraint equations
\beq
\bra{l}
\lb{act}
{\cal H} |n_{\alpha ,i_{\alpha }}
\rangle _{(1\leq \alpha \leq N)}
=E\ |n_{\alpha ,i_{\alpha }}
\rangle _{(1\leq \alpha \leq N)}
\\
{\bf P} |n_{\alpha ,i_{\alpha }}
\rangle _{(1\leq \alpha \leq N)} =
{\Pi _{\alpha =1}^{N}}\
\left( {\Pi _{i_{\alpha }=1}^{k}}\ \
\eta _{\alpha}^{i_{\alpha }-1}\right)
|n_{\alpha ,i_{\alpha }}
\rangle _{(1\leq \alpha\leq N)}\; 
\era
\eeq
where $E$ is the energy spectrum
and ${\Pi _{\alpha =1}^{N}}
\left( {\Pi _{i_{\alpha }=1}^{k}}\ \
\eta _{\alpha }^{i_{\alpha }-1}\right) $
should be equal to the unity in order
to ensure ${\Z}_{k}$ invariance.
To do so, let us first recall some useful features for our
computation. Since for a fixed particle
$\alpha $, the ${\Z}_{k}$
symmetry still commutes with the one electron
Hamiltonian ${\cal H}_{\alpha }$;
i.e.
\beq
{\cal H}_{\alpha }\
{\bf P}={\bf P}\ {\cal H}_{\alpha }
\eeq
the one particle vacuum state is also degenerate
$|{\bf v_{\alpha}}>\equiv\{|\alpha,i_{\alpha}>,\;
i_{\al}=1,\cdots,k\}$.
It is a $k$-dimensional vector with the
following properties:
\beq
\bra{l}
{\bf Q}|\alpha, i_{\alpha }>=|\alpha, i_{\alpha}+1>
\\
{\bf P}|\alpha, i_{\alpha }>=\eta _{\alpha }^{i_{\alpha }}
|\alpha, i_{\alpha }>\; 
\era
\eeq
with $\eta _{\alpha }^{k}=1$, and
\beq
\bra{l}
a_{\alpha i_{\alpha }}^{\pm }
|\alpha, i_{\alpha }>=c_{\alpha }^{\pm }
|\alpha, i_{\alpha }+1>\\
c_{\alpha }^{+}\ |\alpha, i_{\alpha }>=0
\\
H_{\alpha }\ |\alpha, i_{\alpha }>
=\frac{\omega }{2}\ |\alpha, i_{\alpha }>.
\era
\eeq
Moreover it is completely reducible, that is
$|{\bf v_{\alpha }}>\ =\oplus_{i_{\alpha }=1}^{k}\
|\alpha, i_{\alpha }>$,
and so one has $k$ identical copies
rotated among each others by the ${\bf Q}$ operator.
Furthermore as the
energy spectrum of the ${\cal H}_{\alpha }$
operator is
\beq
E_{n_{\alpha,i_{\alpha}}}={\omega\ov 2}
\Big(2n_{\alpha,i_{\alpha }}+ 1 \Big)\; 
\eeq
the excited states $|n_{\alpha ,i_{\alpha }}\rangle $
of ${\cal H}_{\alpha }$ satisfying
\beq
{\cal H}_{\alpha }|n_{\alpha ,i_{\alpha }}\rangle =
{\omega\ov 2}\Big(2n_{\alpha,i_{\alpha }}+1\Big)
|n_{\alpha ,i_{\alpha }}\rangle
\eeq
are given by
\beq
|n_{\alpha, i_{\alpha }}\rangle =
\sqrt{\frac{1}{{(n_{\alpha, i_{\alpha }})!}}}
\Big({a_{\alpha i_{\alpha }}^{+}}
\Big)^{{n_{\alpha ,i_{\alpha }}}}
|\alpha, i_{\alpha}\rangle.
\eeq
On the other hand, since
\beq
L_{0}|\alpha, i_{\alpha }>=i_{\alpha } |\alpha, i_{\alpha }>\; 
\eeq
it follows from covariance
under the ${\Z}_{k}$ symmetry that the
$n_{\alpha, i_{\alpha }}$
integers should be such that
\begin{equation}
\lb{cova}
{n_{\alpha ,i_{\alpha }}}=k\ p_{\alpha }+\ i_{\alpha }
\end{equation}
where $p_{\alpha }$ is a positive integer.
As a first result, if one
considers the particular case where
$i_{\alpha }=1$ for all $\alpha $, that
is, for the ${\Z}_{k}$ invariant
vacuum $|v\rangle =|0,1>$,
then the composite following from~(\ref{stat}) is
\beq
\Big(A_{\alpha }^{+}\Big)_{(1)}=a_{\alpha ,k}^{+}
a_{\alpha ,k-1}^{+} \dots a_{\alpha ,1}^{+}.
\eeq
This is the creation operator of the one electron
state
\beq
\Big(|e_{\alpha }^{-}\rangle \Big)_{(1)}=
\Big(A_{\alpha }^{+}\Big)_{(1)} |0,1>
\eeq
with energy
\beq
{\om\ov 2} \Big(2k+1\Big)
\eeq
and position $z_{\alpha }$. More generally
using~(\ref{prop}), one can build $k$
similar one electron states
\beq
\Big(|e_{\alpha }^{-}\rangle \Big)_{(j)}=
\Big(A_{\alpha}^{+}\Big)_{(j)}|v\rangle =|0>\otimes|j>
\eeq
with the same quantum numbers,
by using the $|0>\otimes|j>
\equiv |0,j>$ vacua.

A careful inspection of~(\ref{stat})
reveals that because of the identities~(\ref{prop}),
the expression of $\Big(A_{\alpha }^{+}\Big)_{(j)}$
is in fact $SU(k)$ invariant.
The point is that out of the $a_{\alpha i_{\alpha }}^{\pm }$
operators, one can construct
the following $SU(k)$ invariant condensate
\begin{equation}
\lb{inva1}
A_{\alpha }^{\pm }=\varepsilon ^{i_{1}\dots i_{k}}
\ a_{\alpha i_{1}}^{\pm}\dots a_{\alpha i_{k}}^{\pm }\; 
\end{equation}
where $\varepsilon^{i_{1}\dots i_{k}}$
is the usual completely
antisymmetric $k$-dimensional invariant
tensor. Due to the relations~(\ref{prop})
only the cyclic $k$-terms of the expansion are
non-zero. The terms that
survive, after using ~(\ref{prop}),
depend on the basis vector on which $A_{\alpha}^{\pm }$
acts. Using the notation
\beq
A_{\alpha j}^{\pm }=<j|A_{\alpha }^{\pm }|j>
\eeq
one can rewrite the above relation as
\begin{equation}
\lb{inva2}
A_{\alpha }^{\pm }=\sum_{j=1}^{k}\
A_{\alpha j}^{+} \pi _{j}
\end{equation}
where the $\pi _{j}$'s are the projectors
on the states $|j\rangle $
introduced earlier. This decomposition
in terms of projectors reflects the
property of the resolution of the singularity
by NC geometry. Each component
may be used to a build solution of the
constraint equations~(\ref{ncon}). But before presenting
these solutions, let us make the contact with the result
of~\cite{hellerman1}. Though the
authors of that work have not addressed
the question of the resolution of
singularity by NC geometry, one can still
recover their wavefunction by
considering the special invariant operator
\beq
\lb{siop}
{\rm{Tr}}A_{\alpha}^{+}=
\sum_{j=1}^{k}<j|A_{\alpha }^{+}|j>.
\eeq
With the aid of this operator and the
realization~(\ref{real}), one can check
that the following wavefunction coincides
with that derived in~\cite{hellerman1}
\begin{equation}
\lb{jsgwf}
|\Phi _{v}^{\nu =1/k}\rangle =
{\cal N} \varepsilon ^{\alpha _{1}\dots
\alpha _{N}}\Big({\rm{Tr}}{A_{\alpha _{1}}^{+}}\Big)
\Big({\rm{Tr}}A_{\alpha _{2}}^{+}\Big)^{2}\dots
\Big({\rm{Tr}}A_{\alpha _{N}}^{+}\Big)^{N}\ |0\rangle.
\end{equation}
This, however, is a special solution where
the effects of NC geometry have
been integrated out. To obtain the
wavefunctions for the system of
$N$ electrons with filling factor
$\nu =\frac{1}{k}$, where NC geometry enters in
the game, one should consider the
$A_{\alpha j}^{\pm }$ operators and the
$|0,j\rangle $ vacuum vector instead of scalars
${\rm{Tr}}{A_{\alpha}^{+}}$ and
$|0\rangle $. Since there are $k$
operators $A_{\alpha j}^{\pm }$ and $k$
vacua $|0, j\rangle $, the wavefunctions
$|\Phi _{v}^{\nu =1/k}\rangle $ of
the system fractionate into $k$ irreducible
components as
\beq
\lb{wfsy}
|\Phi _{v}^{\nu=1/k}\rangle = \sum_{j=1}^{k}
\Big(|\Phi _{v}^{\nu =1/k}\rangle \Big)_{(j)} \pi_{j}
\eeq
where each component
$\Big(|\Phi _{v}^{\nu =1/k}\rangle \Big)_{(j)}$
describes a vacuum configuration given by
\begin{equation}
\lb{vcon}
\Big(|\Phi _{v}^{\nu =1/k}\rangle \Big)_{(j)}=
{\cal N} \varepsilon ^{\alpha_{1}\dots
\alpha _{N}}\Big({A_{\alpha _{1}}^{+}}\Big)_{(j)}
\Big(A_{\alpha_{2}}^{+}\Big)_{(j)}^{2}\dots
\Big(A_{\alpha _{N}}^{+}\Big)_{(j)}^{N}\ |0,j\rangle\; 
\end{equation}
with ${\cal N}$ a normalization factor
and $\varepsilon ^{\alpha_{1}\dots \alpha _{N}}$ is
the usual completely $SU(N)$ invariant tensor.
These degenerate solutions are
rotated under ${\Z}_{k}$
automorphisms and form all
together a cycle of $k$ vertices
in one-to-one correspondence with
the $k$ one-dimensional irreducible representations
of ${\Z}_{k}$ (see Figure 1). This is a remarkable
result which should be understood as a
consequence of NC geometry which acts by
lifting $SU(k)$ singularity of the
Laughlin wavefunctions. In the above relation,
the $\Big(A_{\alpha_{n}}^{+}\Big)^{n} $ operator may be
interpreted as the operator of creation of
an electron at the position $z_{\alpha _{n}}$
with an energy $kn$ in the units of the frequency.
The energy of the above vacuum configuration~(\ref{vcon})
in units of the $\omega $ frequency is
\begin{equation}
\lb{ener}
E_{v}=\frac{N}{2}\Big( kN + k + 1 \Big)\; .
\eeq
It behaves as
\beq
\lb{ener1}
E_{v}\sim \frac{k}{2}N^{2}
\eeq
for large $N$ and agrees with
the expression given in~\cite{hellerman1}.
We will turn to this behaviour in section~4 when
we study FQH states that are not of the
Laughlin type.

We end this subsection by noting that
the wavefunctions associated with~(\ref{jsgwf},\ref{wfsy})
may be derived from the ``non-commutative'' extension of the
Laughlin wavefunctions~(\ref{lau}).
Replacing the lower case (commutative) $z_{\alpha }$
variables by their non-commutative analogues $Z_{\alpha }$, one
gets the following generalized wavefunctions
\beq
\Psi _{L}^{NC}=\Pi _{{\alpha }<{\beta }
=1}^{N}\Big(Z_{\alpha }-Z_{\beta }\Big)^{k}\exp
\left( -\frac{B}{4}\sum_{{\rho =1}}^{N}Z{_{\rho }}
Z{_{\rho }^{+}}\right).
\eeq
Now using the realization
$Z_{\alpha }=W_{\alpha }\otimes {\bf Q}$,
as well as the algebraic relations ${\bf Q}^{k}={\bf I}$
and ${\bf Q}^{\dagger }={\bf Q}^{k-1}$, one
sees that the monomials $\Big(Z_{\alpha }-Z_{\beta }\Big)^{k}$
and the quadratic
objects $Z{_{\rho }}Z{_{\rho }^{+}}$
are in the centre of the $\mathcal{D}%
\left({\Z}_{k}\right) $ representation;
that is
\beq
\Big(Z_{\alpha}-Z_{\beta }\Big)^{k}=\Big(W_{\alpha }-
W_{\beta }\Big)^{k}\otimes {\bf I}
\eeq
and
\beq
Z_{\rho } Z_{\rho }^{+}= W_{\rho }
{\bar{W}}_{\rho } \otimes {\bf I}
\eeq
are proportional to the identity. As such
the above matrix wavefunctions split,
as a sum over the projectors on the
$\mathcal{D}\left({\Z}_{k}\right)$
representation states, as
\beq
\Psi_{L}^{NC}=\sum_{j=1}^{k}<j|\Psi _{L}|j>\pi _{j}.
\eeq
This result coincides exactly with the
expressions~(\ref{jsgwf},\ref{wfsy}) derived
by using the $a^{\pm }$ operator analysis.

\begin{figure}
\begin{center}
\includegraphics{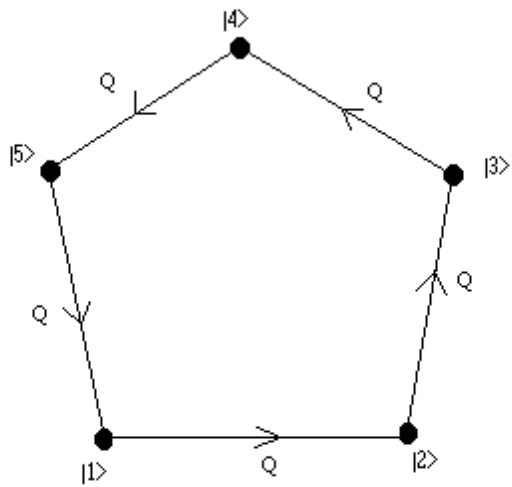}
\end{center}
\caption{{\protect\small {{{\it The vertices of this polygon
represents the $k$ Laughlin wavefunctions (here we have taken a
pentagon with $ k=5$ vertices). The dots are associated with the
$k$ characters of {${\Z}_{k}$}}\ {\it and are rotated under
${\bf Q}$}\ {\it automorphisms as shown on the figure. The
factorisation of the ${\Z}_{k}$}\ {\it invariance implies the
shrinking of the $ k$ vertices down to $j=k$ and as a consequence
one recovers the usual degenerate expression obtained
in}~\cite{hellerman1}.}}}}
\end{figure}

\subsection{Non--commutative matrix model}%----------------------------------

In this section we want to extend the previous
results, especially
those in connection with discrete symmetries
and NC geometry for the matrix
model formulation~\cite{susskind, poly} of
the Laughlin states $\nu ={1\ov k}$ . In this
formulation, the $N$ classical particles are
roughly speaking described by
the $Z_{\alpha \alpha }$ diagonal entries
of a $N\times N$ matrix $Z$
while their ``mutual interactions'' are carried
by the non-diagonal terms
$Z_{\alpha \beta }, \alpha \neq \beta $.
The corresponding creation and
annihilation operators $a_{\alpha \beta }^{\pm }$
of the quantum system are
valued in ${\rm{\bf Adj}}(U\left( N\right))$, contrary to
the previous study where they
were in the ${\bf N}$ and ${\bf{\bar N}}$
representations. We will
give here below a correspondence rule allowing
one to obtain the spectrum of the
matrix model just from the results of the
analysis of subsection (2.1). This
derivation allows us to discover another
remarkable property of the
Polychronachos field operator
and shows that this field operator is just
the leading one of a more general situation
to be considered in section 4.

For a system with a finite number $N$ of
electrons, the action of the matrix
model reads, in terms of the $Z$
and ${\bar Z}$ dynamical variables, as
\begin{equation}
\lb{action}
S={\frac{k}{4\theta }}\int dt\;
{\rm{Tr}}\left( i{\bar{Z}}DZ-\omega Z{\bar{Z}}\right) +
\frac{i}{2}\int dt\; \Psi ^{\dagger }
D\Psi +\frac{k}{2}\int dt\; {\rm{Tr}}A+ hc.
\end{equation}
The matrix variables involved in this
Lagrange description are: (i) A
complex $(0+1)D$ field $Z(t)$
consisting of two hermitian $N\times N$ matrix
fields: $X_{ij}^{1}$ and $X_{ij}^{2}$
transforming in the $U(N)$ adjoint
representation ${\bf N}\otimes {\bf {\bar{N}}}$.
(ii) The Polychronakos field
$\Psi $ in the $U(N)$ fundamental representation
${\bf N}$ and (iii) the
Lagrange matrix field $A_{0}$
transforming in ${\bf N}\otimes {\bf {\bar{N}}}
$ and carrying the constraint of the system.
In addition to the usual
$\left( Z-Z\right) $, $\left( Z-A-Z\right) $,
and $\left( \Psi -A\right) $
couplings, there is moreover a harmonic
oscillator potential term $\left(\omega Z{\bar{Z}}\right) $
serving to glue together the electrons on a disc
forming then a droplet system of radius
$R\sim \sqrt{\frac{2\left(k+1\right) N}{B}}$
and an area~\cite{hellerman2}
\beq
2\pi R^{2} = {2\pi\ov N}{\rm{Tr}}Z{\bar{Z}}.
\eeq
The above action has a one-dimensional
$U(N)=U(1)\times SU(N)$ gauge
invariance which can be used to fix the extra
non-physical degrees of
freedom involved in the above action.
The presence of the term $\frac{k}{2}
\int dt\; {\rm{Tr}}A$ shows
that~(\ref{action}) is actually a constrained
system. This constraint equation, which reads as
\begin{equation}
\lbrack Z,{\bar{Z}}]_{mn}+\frac{\theta }{2k}
\Psi _{m}\Psi _{n}^{+}=\frac{1}{2}
\theta \delta _{mn}\; 
\end{equation}
requires that the $\left( N^{2}-1\right)$ charge
operators of the $SU\left(N\right) \subset U(N)$
gauge invariance are constrained to zero, while the
$U(1)$ charge is fixed to the value $Nk$ as shown
in~(\ref{ncon}).

Introducing the $N^{2}$ creation and
$N^{2} $ annihilation operators
\beq
\bra{l}
b_{\alpha \beta }^{+}\equiv \sqrt{\frac{k}
{2\theta }}(X_{\alpha \beta }+iY_{\alpha \beta })
=\sqrt{\frac{k}{2\theta }}
Z_{\alpha \beta }, \ \
(N\; {\it operators}\; \Psi _{\alpha }^{+})\\
b_{\alpha \beta }^{-}\equiv 
\sqrt{\frac{k}
{2\theta }}(X_{\alpha \beta }-iY_{\alpha \beta })=
\sqrt{\frac{k}
{2\theta }}{\bar{Z}}_{\alpha \beta },\ \
(N\; {\it{operators}}\;
\Psi _{\alpha }^{-})
\era
\eeq
one can derive the quantum spectrum of
the action~(\ref{action}) by following the
same lines of argument we have made earlier. The same
results may be also derived
directly by using the following correspondence
rule:
(i) Insert NC geometry effects by introducing
the internal structure which
allows the replacement of $b_{\alpha \beta }^{\pm }$
by the more general ones
$\left( {\bf b}^{\pm }\right)_{\alpha
i_{\alpha }}^{\bar{\alpha }
i_{\bar{\alpha }}}$
(ii) Associate with the operators
$a_{\alpha i_{\alpha}}^{+}$
of subsection~(2.1), transforming as
$\left( {\bf N},{\bf k}\right)$
under $SU(N)\otimes SU(k)$, the two following operators
$\left( {\bf b}^{+}\right)_{\alpha i_{\alpha }}^{\bar{\alpha }
i_{\bar{\alpha }}}$
and ${\bf \psi }_{\alpha i_{\alpha }}^{+}$
transforming respectively as
$\Big({\bf adj}(U(N)),{\bf adj}(U(k))\Big) $
and $\left( {\bf N},{\bf k}\right).$
Here the ${\bf \psi }_{\alpha i_{\alpha }}^{+}$'s are the
creation operators associated with the $\Psi $ field
and the annihilation
operators are in the complex conjugate representations.
The novelty here is
that the creation and annihilation associated
to the $Z$ and $\bar{Z}$
matrix variables carry $\left( 2+2\right) $
indices. The $\Psi $
representation turns out to be the field one
needs to reduce by contraction
these indices down to a $\left( {\bf N},{\bf k}\right) $
representation as
shown on the following relation
\begin{equation}
\lb{repr}
a_{\alpha i_{\alpha }}^{+}\ \
\lga \ \ \ {\bf a}_{\alpha i_{\alpha}}^{+}=
\left( {\bf \psi }^{+}\cdot
{\bf b}^{+}\right) _{\alpha i_{\alpha}}
\eeq
where
\beq
\left( {\bf \psi }^{+}\cdot
{\bf b}^{+}\right) _{\alpha i_{\alpha}}
=\sum_{\beta =\bar{\beta }=1}^{N}
\sum_{j_{\beta}=1}^{k}{\bf \psi }_{\beta j_{\beta }}^{+}
{\bf b}_{\alpha i_{\alpha }}^{+\bar{\beta }
j_{\bar{\beta }}}.
\end{equation}
This ${\bf a}_{\alpha i_{\alpha }}^{+}$\
operator carries one energy
excitation $\left( {\bf b}^{+}\right) $
and one $U\left( 1\right) $ charge since
\beq
\left[ J_{0},{\bf a}_{\alpha
i_{\alpha }}^{+}\right] =+{\bf a}
_{\alpha i_{\alpha }}^{+}
\eeq
where $J_{0}$\ is the operator counting the
number ${\bf \psi }^{+}$'s as shown
in~(\ref{ncon}). A similar result is also valid
for the $A_{\alpha j}^{\pm }\ \pi _{j}$
composite operators~(\ref{inva1},\ref{inva2}). In
this case, the correspondence rule is
\begin{equation}
\lb{rule}
A_{\alpha }^{+}=\sum_{l=1}^{k}A_{\alpha l}^{+}
\pi _{l}\ \ \ \rightarrow
\ \ {\bf A}_{\alpha }^{+}=
\sum_{l=1}^{k}{\bf A}_{\alpha l}^{+} \pi_{l}
\eeq
and
\beq
\sum_{l=1}^{k}{\bf A}_{\alpha l}^{+} \pi_{l}
=\sum_{l=1}^{k} <l| \varepsilon ^{i_{1}
\dots i_{k}}
{\bf a}_{\alpha i_{1}}^{\pm }
\dots {\bf a}_{\alpha i_{k}}^{\pm } |l> \pi _{l}
\end{equation}
where the ${\bf a}_{\alpha i_{j}}^{\pm }$'s,
are given in~(\ref{repr})
and the $\pi _{j}$'s\ are the
projectors on the states $|j\rangle $
introduced earlier. The ${\bf A}_{\alpha l}^{+}$
operators carry $k$ energy excitation units and $k$ charges
of $U\left( 1\right) $; i.e.
\beq
\left[ J_{0},{\bf A}_{\alpha l}^{+}\right] =+k
{\bf A}_{\alpha l}^{+}.
\eeq
Note in passing that~(\ref{repr},\ref{rule}) are not the unique
way to get condensate representation from the
$\left( {\bf b}^{+}
\right)_{\alpha i_{\alpha }}^{\bar{\alpha }
i_{\bar{\alpha }}}$'s and the
${\bf \psi }_{\alpha i_{\alpha }}^{+}$'s.
One may also define other classes
of $SU\left( k\right) $ invariants in
an analogous way to what we have done for~(\ref{repr}).
For instance, one can define the two following condensates
\beq
\bra{l}
\Psi_{\alpha }^{\pm }=
\sum_{l=1}^{k}<l|\Psi _{\alpha }^{\pm }|l>\pi _{l}
\\
<l|\Psi _{\alpha }^{\pm }|l>=\varepsilon^{i_{1}\dots i_{l}}
{\bf \psi }_{\alpha i_{1}}^{\pm }
\dots {\bf \psi }_{\alpha i_{l}}^{\pm }
\era
\eeq
and
\beq
\bra{l}
\left({\bf B}^{\pm }\right) _{\alpha }^{\bar{\alpha }}=
\sum_{l_{1},l_{2}=1}^{k}<l_{1},l_{2}|
\left( {\bf B}^{\pm }\right) _{\alpha}^{\bar{\alpha }}
|l_{1},l_{2}>\pi _{l_{1}l_{2}} \\
<l_{1},l_{2}|\left( {\bf B}^{\pm }
\right) _{\alpha }^{\bar{\alpha }}
|l_{1},l_{2}>=\varepsilon^{i_{1}
\dots i_{l_{1}}}\ \varepsilon _{j_{1}\dots
j_{l_{2}}} \left( {\bf b}^{\pm }
\right) _{\alpha i_{1}}^{\bar{\alpha }
j_{1}}\dots \left( {\bf b}^{\pm }
\right) _{\alpha i_{l_{1}}}^{\bar{\alpha }
j_{l_{2}}}
\era
\eeq
where $\pi _{ij}=\pi _{i}\otimes \pi _{j}$. The
$\Psi_{\alpha }^{+}$ operators carry $k$ charges of
$U\left( 1\right) $ since
\beq
\left[ J_{0},\Psi _{\alpha }^{+}\right]
=+k \Psi _{\alpha }^{+}\; 
\eeq
while
$\left( {\bf B}^{\pm }\right)_{\alpha }^{\bar{\alpha }}$
carry $k$ units of the energy excitations.
In terms of these operators and following
the same philosophy as before, we can build
an object
\beq
{\bf E}_{\alpha }^{+}=\left( \Psi ^{+}
\cdot {\bf B}^{+}\right) _{\alpha }
\eeq
where
\beq
\left( \Psi ^{+}
\cdot {\bf B}^{+}\right) _{\alpha }=
\sum_{\beta =1}^{N}\Psi _{\beta }^{+}
\left( {\bf B}^{+}\right) _{\alpha }^{\bar{\beta }}\; 
\eeq
carrying $k$ charges of $U\left( 1\right) $ and $k$
energy excitation units, exactly as for the composite
${\bf A}_{\alpha l}^{+}$ of~(\ref{rule}).
In fact due to the factorisation~(\ref{real}),
the ${\bf A}_{\alpha }^{+}$ and
${\bf E}_{\alpha }^{+}$\ are proportional and then we
will use the ${\bf A}_{\alpha }^{+}$ objects.

One can also deduce the Hamiltonian
${\cal H}$ associated with the matrix
model by using~(\ref{rema}). It reads in terms of the
$\left( {\bf a}^{\pm}\right) _{\alpha
i_{\alpha }}^{\beta j_{\beta }}$'s as
\begin{equation}
{\cal H}={\omega\ov 2} \Big(2{\cal N}_{b}+N^2 \Big)
\end{equation}
where ${\cal N}_{b}$ is the number operator counting
$\left({\bf b}^{\pm}\right)_{\alpha }^{\bar{\alpha }}.
\left( {\bf Q}^{\pm 1}\otimes {\bf Q}^{\pm 1}\right) $.
It is proportional to the sum over the $N^{2}$ number operator
$\left({\bf b}^{+}\right)_{\alpha}^{\bar{\alpha }}
\left( {\bf b}^{-}\right)_{\alpha}^{\bar{\alpha}}$
times the $\left( {\bf I}_{k}\otimes {\bf I}_{k}\right)$
identity operator. The wavefunctions can
also be worked out immediately from~(\ref{vcon});
all one has to do is to replace
the $A_{\alpha }^{+}$ operators of~(\ref{vcon})
by the expressions~(\ref{repr},\ref{rule}).

The above relations may also be derived
by following the standard approach. The
commutation relations for the matrix
model are given by
\beq
\bra{l}
\left[ \left( {\bf b}^{-}\right)_{\alpha
i_{\alpha }}^{\bar{\alpha }i_{\bar{\alpha }}},
\left( {\bf b}^{+}\right) _{\beta j_{\beta }}^{\bar{\beta }
j_{\bar{\beta }}}\right] =\delta _{\alpha \beta}
\delta ^{\bar{\alpha }\bar{\beta }}
\delta _{i_{\alpha }j_{\beta  }}
\delta ^{{{i}}_{\bar\alpha}{{j}}_{\bar\beta}}
\\
\left[ {\bf \psi }_{\alpha i_{\alpha }}^{-},
{\bf \psi}_{\beta j_{\beta }}^{+}\right] =
\delta _{\alpha \beta }\delta _{i_{\alpha }j_{\beta}}
\\
\left[ \left( {\bf b}^{\pm }\right) _{\alpha
i_{\alpha }}^{\bar{\alpha }i_{\bar{\alpha }}},
\left( {\bf b}^{\pm }\right) _{\beta i_{\beta }}^{\bar{\beta }
i_{\bar{\beta }}}\right] =0\\
\left[ \left( {\bf b}^{\pm}\right)_{\alpha
i_{\alpha }}^{\bar{\alpha }
i_{\bar{\alpha }}},
{\bf \psi }_{\beta j_{\beta }}^{\pm }\right]=0\\
\left[ {\bf \psi }_{\alpha i_{\alpha }}^{\pm },
{\bf \psi }_{\beta j_{\beta }}^{\pm }\right] =0.
\era
\eeq
Since $\left( {\bf b}^{\pm }\right)_{\alpha
i_{\alpha }}^{\bar{\alpha }
i_{\bar{\alpha }}}$ and
${\bf \psi }_{\alpha i_{\alpha }}^{\pm }$ may
also be expressed in a condensed form by using
the realization~(\ref{real}) as follows
\beq
\sum_{i_{\alpha },i_{\bar{\alpha }}=1}^{N}
\left( {\bf b}^{\pm }\right)_{\alpha i_{\alpha }}^{\bar{\alpha }
i_{\bar{\alpha }}}
=\left( {\bf b}^{\pm }
\right)_{\alpha }^{\overline{\alpha }}.
\left( {\bf Q}^{\pm 1}\otimes {\bf Q}^{\pm 1}\right)\; 
\eeq
namely
\beq
\sum_{i_{\alpha },i_{\bar{\alpha }}=1}^{N}
\left( {\bf b}^{\pm }\right)_{\alpha i_{\alpha }}^{\bar{\alpha }
i_{\bar{\alpha }}}=
\left( {\bf b}^{\pm }\right) _{\alpha }^{\bar{\alpha }}
\sum_{i_{\alpha },i_{\bar{\alpha }}=1}^{N}
{\bf f}_{i_{\alpha }}\ {\bf f}_{i_{\bar{\alpha }}}
\eeq
and similarly
\beq
\sum_{i_{\alpha }=1}^{N}{\bf \psi }_{\alpha
i_{\alpha }}^{\pm }={\bf \psi }_{\alpha }^{\pm }.
{\bf Q}^{\pm 1}\; 
\eeq
the above commutation relations can be
rewritten as
\beq
\bra{l}
\left[ \left( {\bf b}^{-}\right)_{\alpha }^{\bar{\alpha }},
\left( {\bf b}^{+}\right) _{\beta }^{\bar{\beta }}\right]
=\delta_{\alpha \beta } \delta^{\bar\alpha \bar\beta }\\
\left[ {\bf \psi }_{\alpha }^{-},{\bf \psi }_{\beta }^{+}
\right] =\delta _{\alpha \beta }\\
\left[ \left( {\bf b}^{\pm }
\right)_{\alpha }^{\bar{\alpha }},
\left( {\bf b}^{\pm }
\right) _{\beta }^{\bar{\beta }}\right] =0\\
\left[ \left( {\bf b}^{\pm }
\right) _{\alpha }^{\bar{\alpha }},
{\bf \psi }_{\beta }^{\pm }\right] =0\\
\left[ {\bf \psi }_{\alpha }^{\pm },
{\bf \psi }_{\beta }^{\pm }\right] =0.
\era
\eeq
Here also NC geometry lifts the degeneracy of
the vacuum configurations with minimal
energy
$E_{v}={N\ov 2} \Big( kN + k + 1 \Big)$
and $Nk$ charges of $U\left( 1\right) $.
The vacuum wavefunction of the FQH states at
$\nu =\frac{1}{k}$ is in the
centre of the ${\cal D}\left( {\Z}_{k}\right) $
representation and reads as
\begin{equation}
|\Phi _{v}^{\nu =1/k}\rangle =
{\cal N} \varepsilon ^{\alpha _{1}\dots
\alpha _{N}}({\bf A}{_{\alpha _{1}}^{+}})
\Big({\bf A}_{\alpha_{2}}^{+}\Big)^{2}\dots
\Big({\bf A}_{\alpha _{N}}^{+}\Big)^{N}\ {\bf I}\;
|v\rangle
\end{equation}
where ${\bf I}$ is the identity operator of
${\cal D}\left( {\Z}_{k}\right) $ and $|v\rangle $
the vacuum vector and the ${\bf A}{_{\alpha j}^{+}}$'s
are as in~(\ref{rule}). Viewed from the base of fibration
${\RR}_{\theta }^{2}\times {\cal D}\left( {\Z}_{k}\right) $,
this relation reduces to the Hellermann and Van Raamsdonk
(HR) wavefunction $|\Phi_{HR}>$ obtained
in~\cite{hellerman1} and which reads, in terms of our
notational convention, as
\beq
|\Phi _{HR}^{\nu =\frac{1}{k}}>={\cal N}
\varepsilon^{\alpha _{1}\dots \alpha _{N}}
\Big(\Psi ^{+}.{C_{\alpha _{1}}^{+}}\Big) \Big(\Psi^{+}.
\Big(C_{\alpha _{2}}^{+}\Big)^{2}\Big)\dots
\Big(\Psi ^{+}.\Big(C_{\alpha _{N}}^{+}\Big)^{N}\Big)
|0\rangle.
\eeq
Recall that in this representation, the vacuum $|0>$ ignores
all aspects of the ${\Z}_{k}$ symmetry of the Laughlin wavefunctions.

%%%%%%%%%%%%%%%%%%%%%%%%%%%%%%%%%%%%%%%%%%%%%%%%%%%%%%%%
\section{$\nu_{k_1k_2}$ Fractional quantum Hall states}
%%%%%%%%%%%%%%%%%%%%%%%%%%%%%%%%%%%%%%%%%%%%%%%%%%%%%%%

Although the $\nu =\frac{2}{5}$ FQH state is not
of the Laughlin type, it shares, however, some basic
features of Laughlin fluids. The point is that
from the standard definition of the filling
factor $\nu =\frac{N}{N_{\phi }}$, the state
$\nu =\frac{2}{5}$ can naively be thought of as
corresponding to $\nu =\frac{N}{N_{\phi }}$
where the number $N_{\phi }$ of flux quanta is given
by a fractional amount of the electron number; that
is
\beq
N_{\phi }=(3-{\frac{1}{2}})N.
\eeq
In fact this way of viewing
things reflects just the original idea of the
hierarchical construction of FQH
states of general filling factor $\frac{p}{q}$,
considered years ago by many
FQH authors. In Haldane's hierarchy~\cite{haldane}
construction, for instance, the
$ K_{IJ}$ matrix of~(\ref{csg}) is
taken as
\[
K_{IJ}=\left(
\begin{array}{cccccc}
p_{1} & -1 &  &  &  &  \\
-1 & p_{2} & -1 &  &  &  \\
& -1 & . & . &  &  \\
&  & . & . & . &  \\
&  &  & . & . & -1 \\
&  &  &  & -1 & p_{n}
\end{array}
\right)
\]
with $p_{1}$\ an odd integer and the others
$p_{i}$'s even. The filling
factor is given by the continuous fraction
\beq
\nu_{p_{1}...p_{n}}=
\frac{1}{p_{1}-\frac{1}{p_{2-...}}}.
\eeq
For the level two
of the hierarchy $\left( n=2\right) $,
the elements of the series
\beq
\nu_{p_{1}p_{2}}=\frac{p_{2}}{p_{1}p_{2}-1}
\eeq
correspond to taking $N_{\phi }$
as given by a specific rational
factor of the electron number; i.e.
\beq
N_{\phi}=(p_{1}-\frac{1}{p_{2}})N.
\eeq
Upon setting
\beq
\bra{l}
k_{1}=p_{1}\\
k_{2}=k_{1}(k_{1}p_{2}-1)
\era
\eeq
the rational
factor $\frac{p_{1}p_{2}-1}{p_{2}}$
can be brought into the following
suggestive form $\frac{k_{1}k_{2}}
{ k_{1}+k_{2}}$, and so the filling factor
$\nu _{p_{1} p_{2}}\equiv \nu_{k_{1} k_{2}}$
splits as
%~\cite{saidi3}
\begin{equation}
\nu _{k_{1} k_{2}}=\frac{1}{k_{1}}+\frac{1}{k_{2}},
\ \ \ \ \ k_{2}>k_{1}.
\end{equation}
Therefore FQH states with $\nu _{k_{1} k_{2}}$
may, under some conditions, be thought
of as consisting of two coupled
Laughlin states of filling factors
$\nu _{k_{1}}=\frac{1}{k_{1}}$ and
$\nu_{k_{2}}=\frac{1}{k_{2}}$ respectively.
The choice of $p_{1}$ an odd integer
and $p_{2}$ even ensures automatically
that both $k_{1}$ and $k_2$ are odd
integers, and so both of the
$\nu _{k_{i}}=\frac{1}{k_{i}}$
$(i=1,2)$ FQH branches
describe fermions. This feature, which is valid
for any level $n$ of the
Haldane hierarchy, reads, for the special
$\nu =\frac{2}{5}$ example we will
be considering to illustrate our results, as
\begin{equation}
\nu =\frac{2}{5}\equiv\frac{1}{3}+\frac{1}{15}.
\end{equation}
To study vacuum configurations of such
FQH states, it is interesting to fix
some terminology and specify the hypothesis
we will be using. As far as
terminology and notational convention
are concerned, let $N_{1}$
(resp. $ N_{\phi _{1}}$) be the number of
electrons (resp. quantum flux) in the $\nu_{k_1}=
\frac{1}{k_{1}}$ FQH fundamental state and
$N_{2}$ (resp. $N_{\phi _{2}}$) be the number
of electrons (resp. quantum flux) in $\nu _{2}=\frac{1}{k_{2}}$.
From the relation $\nu _{k_{i}}=\frac{1}{k_{i}}$,
we have the identities $ N_{\phi _{i}}=k_{i}N_{i}$.
Let also $B\equiv B_{1}$ be the external magnetic
field viewed by the $N_{1}$ electrons of
the $\nu _{k_{1}}$
FQH fundamental state and $B^{\ast }\equiv B_{2}$
be the effective magnetic
field felt by the $N_{2}$ electrons of
the $\nu _{k_{2}}$
hierarchical state. The relation between the
$B$ and $B^{\ast }$ fields is
\beq
\frac{B_{2}}{B_{1}}=\frac{k_{2}}{k_{1}}\equiv
\left( k_{1}p_{2}-1\right).
\eeq
For the $\nu =\frac{2}{5}$ state, we have the
relation $B_{2}=5B_{1}$. A way to
derive this result is to use the following
semi-classical analysis. First
denote by $\left\{ Z_{1,\alpha },{\bar{Z}}_{1,\alpha };
1\leq \alpha \leq
N_{1}\right\} $ the quantum coordinate
operators associated with the $\nu_{k_{1}}=
\frac{1}{k_{1}}$ state and by
$\left\{ Z_{2,a},{\bar{Z}}_{2,a};
1\leq a\leq N_{2}\right\} $ those associated with
the $\nu _{k_{2}}=\frac{1}{k_{2}}$ state.
The dynamics of these two sets of
matrix variables is given by actions of
type~(\ref{action}).
Then use the following
quantum constraint equations
by treating for the moment the two
Laughlin states $\nu _{k_{1}}$\ and
$\nu _{k_{2}}$\ as independent:
\beq
\bra{l}
\lb{3ncon}
\left[ Z_{1,\alpha },{\bar{Z}}_{1,\beta }\right]
= i\theta _{1}\delta_{\alpha \beta } \\
\left[ Z_{2,a},{\bar{Z}}_{2,b}\right]  =
 i\theta _{2}\delta _{ab}\; 
\era
\eeq
where, according to the result of
Susskind, the $\theta _{j}$ parameters
in the large $N$ limits are given by
\beq
\theta _{j}={\rm const}\;\frac{k_{j}}{B_{j}}.
\eeq
Since $\theta _{j}$ is interpreted as
the effective size occupied by an
electron in the quantum space
${\RR}_{\theta }^{2}$, it follows from the
indiscernability hypothesis that
$\theta _{1}=\theta _{2}$ and consequently
\beq
\lb{equiv}
\frac{k_{1}}{B_{1}}=\frac{k_{2}}{B_{2}}.
\eeq
Setting $B_{i}=l_{i}^{-2}$ where $l_{1}$ and $l_{2}$
are the so-called magnetic lengths associated to $B_{1}$
and $B_{2}$ respectively, the previous relation then reads as
\begin{equation}
\lb{n3ncon}
k_{i}l_{i}^{2}={\rm const}.
\end{equation}
As we are dealing with $N=\left( N_{1}+N_{2}\right)$
electrons coming from
two origins and since $k_{2}>k_{1}$,
it is helpful to introduce the following useful
terminology. Thus we will refer
to the elementary flux $\phi _{1}$
occupying the area ${\cal A}_{1}\sim
\frac{1}{B_{1}}=l_{1}^{2}$ in the $\nu_{k_{1}}$
state as {\it quasi-electrons}
and to those elementary
fluxes $\phi _{2}$, of size ${\cal A}_{2}\sim
\frac{1}{B_{2}}=l_{2}^{2}$, in
the $\nu_{k_{2}}$ state as
{\it quasi-muons}\footnote{{ Muons
$\mu $ are elementary leptons with
properties similar to
electrons, except for being more massive
than the electrons. They are introduced
here purely to simplify the presentation.
Our $\mu $'s are in fact those
electrons coming from the condensation
of the quasi-particles of the
$\nu =\frac{1}{15}$ state}}.
This terminology should not be confused with
various ones used in condensed matter physics
literature. In our case, all
that this appellation means is that
when $k_{1}$ {\it quasi-electrons}
(resp. $k_{2}$ {\it quasi-muons})
condensate, they give rise to one real
electron (resp. one muon by extension).

Using the results of section 2, one can
first write down the fundamental wavefunctions
$|\Phi_{1},v_{k_{1}}\rangle\otimes
|\Phi_{2},v_{k_{2}}\rangle$
for the uncoupled system; that is for
the situation where each of the
filling factors $\nu _{k_1}$
and $\nu _{k_{2}}$
are treated separately. This is the case
for instance of two FQH layers distant
enough so that they cannot feel each other.
However, this is not the case here. The
system in the present study consists
of one layer only and the $\nu _{k_{1}}$
and $\nu _{k_{2}}$
states should be coupled. If
we denote by $a_{\alpha i_{\alpha }}^{+}$
the creation operators of
{\it quasi-electrons} (resp. $d_{ai_{a}}^{+}$
the creation operators of
{\it quasi-muons}) and by $A_{\alpha }^{+}$
the creation operators of one electron
with energy $k_1$ in $\omega_1 $ units and $k_1$
charges of $U\left( 1\right)$~(\ref{inva1},\ref{inva2})
(resp. $D_{a}^{+}$ the creation operator
of muons with energy $k_2$
in $\omega_2$ units and $k_2$ charges of
$U^{\prime }\left( 1\right)$~(\ref{inva1},\ref{inva2})),
then the wavefunction
$|\Phi _{v_{12}}\rangle $ describing the vacuum configuration
of $N=\left( N_{1}+N_{2}\right) $ particles is
(with the hypothesis $N_{1}=rN_{2}$ , $r>1$,
as is usually the case for the level two of
the Haldane hierarchy)
\begin{equation}
\lb{wfun}
|\Phi _{v_{12}}\rangle =\varepsilon^{I_{1}\dots
I_{r}}[{\cal A}^{+}]_{I_{1}}
[{{\cal A}^{+}(}D{^{+})^{1}]}_{I_{2}}
\dots \lbrack {{\cal A}^{+}(}D{^{+})^{r-1}]}_{I_{r}}
|0,(s_{1},s_{2})\rangle
\end{equation}
where $I_{j}$ stands for a multi-index
\beq
I_{j}=(\beta _{jN_{2}},\beta_{jN_{2}+1},
\dots, \beta _{jN_{2}+({N_{2}-1})})
\eeq
indexing the $j$-th block;
the tensor $\varepsilon ^{I_{1}\dots I_{r}}$
is the usual $SU\left(N_{1}\right) $ invariant
antisymmetric tensor expressed in terms of
multi-indices; that is
\beq
\varepsilon^{I_{1}\dots I_{r}}=\varepsilon^{\beta_{1}\dots
\beta _{N_{2}}\dots \beta _{rN_{2}}\ldots
\beta _{(r+1)N_{2}-1}}
\eeq
and where the $\left[ {\cal A}^{+}
{\left( D^{+}\right)^{j-1}}\right]_{I_{j}}$
building blocks are given by
\begin{equation}
\left[ {\cal A}^{+}{\left( D^{+}\right)^{j-1}}\right]_{I_{j}}=
\left( A^{+}{\left( D^{+}\right)^{j-1}}\right) _{\beta _{1}}
\left( \left( A^{+}\right)^{2}\left( D^{+}\right)^{j-1}
\right)_{\beta _{2}}\dots \left( \left(A^{+}\right)^{N_{2}}
\left( D^{+}\right) ^{j-1}\right) _{\beta _{N_{2}}}.
\end{equation}
$|0,(s_{1},s_{2})\rangle $ is the vacuum state
transforming as a $\left({\bf k}_{1}{\bf ,k}_{2}\right) $
vector under ${\Z}_{k_{1}}\otimes {\Z}_{k_{2}}$ symmetry.
The energy of the above vacuum configuration
is directly obtained by computing the energy
of the generic building blocs
$\left[ {\cal A}^{+}
{\left( D^{+}\right)^{j-1}}\right] _{I_{j}}$.
The latter have an energy
contribution of the form
$E_{A^{+}}+E _{D^{+}}$ associated
with the $A^{+}$ and $D^{+}$ operators and respectively
equal to
\beq
\bra{l}
E_{A^{+}}={1\ov 2}k_{1}N_{2}\left(N_{2}+1\right)\\
E_{D^{+}}=(j-1) k_{2}N_{2}.
\era
\eeq
Adding these two energies, one obtains
\begin{equation}
E_{A^{+}}+E_{D^{+}}={1\ov 2}
k_{1}N_{2}(N_{2}+1)+\left(j-1\right)
k_{2}N_{2}.
\end{equation}
Then summing over all allowed values of the index
$j$; i.e. $1\leq j\leq r$,
as required by the expression of the wavefunction
$|\Phi _{v_{12}}\rangle $, by taking into
account the relation $N_{1}=rN_{2}$, one gets
\begin{equation}
E_{A^{+}}+E_{D^{+}}=\frac{N_{1}}{2}
\Big[ k_{1}\left(N_{2}+1\right) +
k_{2}\left(r-1\right) \Big]\; .
\end{equation}
Taking into account the vacuum contribution of
the oscillator which is equal to $\frac{N_{1}+N_{2}}{2}$,
one ends up with the following relation
for the vacuum energy of the interacting configuration
\begin{equation}
E_{v_{12}}^{\nu =1/k_{1}+1/k_{2}}=
\frac{1}{2}\Big\{ k_{1}N_{1}N_{2}+N_{1}\left[
k_{2}\left( r-1\right) +k_{1}+1\right] +N_{2}\Big\}.
\end{equation}
Note that for large values of $N_{1}$
and $N_{2}$, but $\frac{N_{1}}{N_{2}}=
\frac{k_{2}}{k_{1}}$ finite, say
\beq
\bra{l}
N_{1}= rN_{2}\equiv rM\\
k_{2}=rk_{1}\; 
\era
\eeq
the vacuum energy of the configuration~(\ref{wfun})
behaves quadratically in $M$ with a coefficient $\frac{k_{2}}{2}$,
namely
\begin{equation}
E_{v_{12}}^{\nu =\left( 1/k_{1}+1/k_{2}\right) }
\sim \frac{k_2}{2}M^2.
\end{equation}
This energy relation is less than the
total energy $\left( E_{v_{1}}^{\nu=1/k_{1}}+
E_{v_{2}}^{\nu =1/k_{2}}\right) $ of the
decoupled configuration $\left( |\Phi _{1},v_{k_{1}}
\rangle \otimes |\Phi _{2},v_{k_{2}}\rangle\right) $
which also behaves quadratically in $M$
as shown here below; but with a coefficient
$\frac{k_{2}\left(r+1\right)}{2}$
larger than that appearing in presence of interactions.
Indeed, using~(\ref{ener},\ref{ener1}), we obtain
\beq
E_{v_{1}}^{\nu =1/k_{1}}+E_{v_{2}}^{\nu =
1/k_{2}} \sim \frac{k_{1}N_{1}^{2}}{2}+
\frac{k_{2}N_{2}^{2}}{2}
\eeq
leading to
\beq
E_{v_{1}}^{\nu =1/k_{1}}+E_{v_{2}}^{\nu = 1/k_{2}}
\sim
\frac{k_{2}(r+1)}{2}\ M^{2}.
\eeq
Therefore the difference between the energies
$\left( E_{v_{1}}^{\nu=1/k_{1}}+E_{v_{2}}^{\nu =1/k_{2}}\right) $
and $E_{v_{12}}^{\nu =\left(1/k_{1}+1/k_{2}\right) }$
of the decoupled configuration and the interacting one is
\begin{equation}
\lb{dif1}
\left( E_{v_{1}}^{\nu =1/k_{1}}+E_{v_{2}}^{\nu =1/k_{2}}\right)
-E_{v_{12}}^{\nu =\left( 1/k_{1}+1/k_{2}\right) }\sim \frac{k_{2}r}{2}\
M^{2}
\end{equation}
showing that
\beq
\lb{dif2}
\left( E_{v_{1}}^{\nu =1/k_{1}}+E_{v_{2}}^{\nu
=1/k_{2}}\right) \sim \left( r+1\right)
E_{v_{12}}^{\nu =\left(1/k_{1}+1/k_{2}\right) }.
\eeq
For the example of the FQH state at filling
factor $\nu =\frac{2}{5}$, the energy of
the decoupled representation $\frac{2}{5}=\frac{1}{3}+\frac{1}{15}$
reads as
\begin{equation}
\lb{decap}
E_{v_{1}}^{\nu =1/3}+E_{v_{2}}^{\nu =1/15}\sim 45 M^{2}
\end{equation}
while that of the interacting one is
\beq
\lb{cap}
E_{v_{12}}^{\nu =\left( 1/3+1/15\right)}\sim \frac{15}{2}M^{2}.
\eeq
It is obvious to see that equations~(\ref{decap},\ref{cap})
verify the relation~(\ref{dif2}), such that
\beq
\left( E_{v_{1}}^{\nu =1/3}+E_{v_{2}}^{\nu
=1/15}\right) \sim 6
E_{v_{12}}^{\nu =\left(1/3+1/15\right) }.
\eeq

In what follows, we propose a matrix model
to describe such FQH states that
are not of Laughlin type. We will focus
our attention on the ${e}-{\mu}$
system we presented above although
most of our results may be extended
to more general FQH systems.

%%%%%%%%%%%%%%%%%%%%%%%%%%%%%%%%%%%%%%%%%%%%%%%%%%%%%
\section{Matrix model for $({e}-{\mu})$ system}
%%%%%%%%%%%%%%%%%%%%%%%%%%%%%%%%%%%%%%%%%%%%%%%%%%%%%

We start by presenting the variables
of the matrix model for the case
of a FQH droplet of $N=\left( N_{1}+N_{2}\right) $
electrons ($N_{1}$ electrons and $N_{2}$ muons)
with filling factor
$\nu_{k_{1}k_{2}}$.
The integers $k_{1}$ and $k_{2}$ are some specific odd
integers; they are essentially given by the
family of integers, $k_{1}=p_{1}$
and $k_{2}=p_{1}(p_{1}p_{2}-1)$ with
$p_{1}$\ odd and $p_{2}$ even,
appearing in the second level of the
Haldane hierarchy. One of the key ideas
of our description is to think about these
states as consisting of two coupled
branches of filling fractors
$\nu _{k_1}=\frac{1}{k_{1}}$ and $\nu_{k_2} =\frac{1}{k_{2}}$.
The second FQH state with $N_{2}$ muons is built
on top of the $\nu_{k_1}$ state; that is it comes
after the condensation of the $N_{1}$
electrons of the $\nu _{k_1}$
state, viewed, by the way, as the
level one of the hierarchy. The matrix fields
involved in our model are of
three kinds: the first fields, which describe
the set of $N_{1}$ electrons in the presence of the
external magnetic field $B$, are associated with the
$e${\it --sector}, the second type of fields describe
the $\mu${\it --sector} and the third type of
fields carry the $({e}-{\mu})${\it --couplings}.

%%%%%%%%%%%%%%%%%%%%%%%%%%%%%%%%%%%%%%%%%%%%%%%%%%%%%
\subsection{${e}$--Sector}
%%%%%%%%%%%%%%%%%%%%%%%%%%%%%%%%%%%%%%%%%%%%%%%%%%%%%

The matrix variables associated with the
$e${\it --sector} of the $({e}-{\mu})$ system
are supposed to describe the $\nu_{k_1}=\frac{1}{k_{1}}$
branch of the fluid. They are given by the usual triplet
$(Z_{1},\Psi_{1},A_{1})$ appearing in the Susskind-Polychronakos
matrix model and have the following $U(N_{1})$ group structure
\beq
\bra{l}
\lb{gstr}
Z_{1} =(Z_{1})_{\alpha }^{\bar{\alpha}}
\sim {\bf N}_{1}\otimes {\bf {\bar{N}}}_{1} \\
{\bar{Z}}_{1} =({\bar{Z}}_{1})_{\bar{\alpha}}^{\alpha }
\sim {{\bar{{\bf N}}}}_{1}\otimes {\bf N}_{1} \\
\Psi _{1} =(\Psi _{1})_{\alpha }\ \ \sim {\bf N}_{1} \\
{\bar{\Psi}}_{1} =({\bar{\Psi}_{1}})^{\bar{\alpha}}\ \
\sim {\bf {\bar{N}}}_{1}  \\
A_{1} =(A_{1})_{\alpha }^{\bar{\alpha}}
\sim {\bf N}_{1}\otimes {\bf {\bar{N}}}_{1}.
\era
\eeq
The matrix model describing the dynamics
of these fields is given by the
following one-dimensional
$U(N_{1})$ gauge invariant action
\begin{equation}
{\cal S}_{1}={\frac{k_{1}}{4\theta _{1}}}
\int dt\; {\rm{Tr}}\left( i{\bar{Z}_{1}}DZ_{1}-\omega _{1}
{\bar{Z}_{1}}Z_{1}\right) +\frac{i}{2}\int dt\;
{\bar{\Psi}}_{1}D\Psi _{1}+\frac{k_{1}}{2}\int dt\;
{\rm{Tr}}A_{1}+ hc
\end{equation}
while the constraint equations,
which are obtained as usual by computing the
equation of motion of $A_{1}$, read as
\begin{equation}
\lb{4ncon}
\lbrack Z_{1},{\bar{Z}_{1}}]_{\alpha }^{\bar{\alpha}}+
\frac{\theta _{1}}{2k_{1}}\Psi _{\alpha }
{\bar{\Psi}}^{\bar{\alpha}}
=\frac{1}{2}\theta_{1}
\delta _{\alpha }^{\bar{\alpha}}.
\end{equation}
Quantum mechanically, these constraint equations
should be imposed on the Hilbert space ${\bf H}$
of the wavefunctions $|\Phi \rangle $ and so~(\ref{4ncon})
should be thought of as
\beq
\bra{l}
J_{\alpha \bar{\alpha}}^{(1)}|\Phi \rangle =0\\
J_{0}^{(1)}|\Phi \rangle =
\left( k_{1}N_{1}=\frac{N_{1}}{\nu _{1}} \right) |\Phi \rangle
\era
\eeq
where $J_{0}^{(1)}$ and $J_{\alpha \bar{\alpha}}^{(1)}$
are the generators of the $U(N_{1})=U(1)\times SU(N_{1})$
gauge symmetry of the action ${\cal S}_{1}$. Later on we
will give the field realization of these charge
operators; but for the moment let us complete the
presentation of the dynamical variables of the
$(e-\mu)$ system.

%%%%%%%%%%%%%%%%%%%%%%%%%%%%%%%%%%%%%%%%%%%%%%%%%%%%%
\subsection{${\mu}$--Sector}
%%%%%%%%%%%%%%%%%%%%%%%%%%%%%%%%%%%%%%%%%%%%%%%%%%%%%

\no The matrix field variables associated with
the subsystem ${\bf\mu}$ containing $N_{2}$
electrons (muons) are quite similar to those appearing in
the ${e}${\it --sector}. These variables, which describe
the $\nu _{k_2}$ branch of the
$\nu_{k_{1}k_{2}}$ fluid, are given by the triplet
$(Z_{2},\Psi _{2},A_{2})$ valued in $U(N_{2})$ group
representations as shown below
\beq
\bra{l}
\lb{grep}
Z_{2} =(Z_{2})_{a}^{\bar{a}}
\sim {\bf N}_{2}\otimes {\bf {\bar{N}}}_{2}\\
{\bar{Z}}_{2} =({\bar{Z}}_{2})_{\bar{a}}^{a}
\sim {\bf \bar{N}}_{2}\otimes {\bf N}_{2}\\
\Psi _{2} =(\Psi _{2})_{a}\ \ \sim {\bf N}_{2} \\
{\bar{\Psi}}_{2} =({\bar{\Psi}_{2}})^{\bar{a}}\ \
\sim {\bf {\bar{N}}}_{2}\\
A_{2} =(A_{2})_{a}^{\bar{a}}
\sim {\bf N}_{2}\otimes {\bf {\bar{N}}}_{2}.
\era
\eeq
The action ${\cal S}_{2}$ of the matrix model for
this {\it sector} reads as
\begin{equation}
\lb{sec2}
{\cal S}_{2}={\frac{k_{2}}{4\theta _{2}}}\int dt\;
{\rm{Tr}}\left( i{\bar{Z}_{2}}DZ_{2}-\omega _{2}Z_{2}
{\bar{Z}_{2}}\right) +\frac{i}{2}\int dt\;{\bar{\Psi}}_{2}
D\Psi _{2}+\frac{k_{2}}{2}\int dt\; {\rm{Tr}}A_{2}+ hc.
\end{equation}
The constraint equations are naturally given by
\begin{equation}
\lb{24ncon}
\lbrack Z_{2},{\bar{Z}_{2}}]_{a}^{\bar{a}}+
\frac{\theta _{2}}{2k_{2}}
\Psi _{a}{\bar{\Psi}}^{\bar{a}}=
\frac{1}{2}\theta _{2}\delta _{a}^{\bar{a}}.
\end{equation}
At the quantum level, they should be thought of as
\beq
\bra{l}
\lb{q4ncon}
J_{a\bar{a}}^{(2)}|\Phi \rangle =0 \\
J_{0}^{(2)}|\Phi \rangle = \left(
k_{2}N_{2}=\frac{N_{2}}{\nu _{2}} \right) |\Phi \rangle
\era
\eeq
where now $J_{0}^{(2)}$ and $J_{a{\bar{a}}}^{(2)}$
are the generators of the $U(N_{2})=U(1)\times SU(N_{2})$
gauge symmetry of the action ${\cal S}_{2}$.

Note that as far as these two pieces of
the total action ${\cal S=}\left({\cal S}_{1}+{\cal S}_{2}+
{\cal S}_{int}\right) $ of the $(e-\mu)$
system are concerned, the full gauge
symmetry is $U(N_{1})\times U(N_{2})$.
The matrix model variables listed above
transform under this invariance as
\beq
\bra{l}
Z_{i} \rightarrow U_{i}Z_{i}U_{i}^{\dagger } \\
\Psi _{i} \rightarrow U_{i}\Psi _{i} \\
A_{i} \rightarrow U_{i}(A_{i}-\partial _{t})
U_{i}^{\dagger }
\era
\eeq
where $i=1,2$ and the $U_{i}$ gauge transformations
are given by
\beq
U_{i}=\exp \left( i\sum_{n=1}^{N_{i}}
T_{n}\lambda _{i}^{n} (t) \right)
\eeq
with $T_{n}$ being the group generators and
$\lambda _{i}^{n}\left( t\right) $ the gauge
parameters.

%%%%%%%%%%%%%%%%%%%%%%%%%%%%%%%%%%%%%%%%%%%%%%%%%%%%%
\subsection{$(e-\mu)$--Couplings}
%%%%%%%%%%%%%%%%%%%%%%%%%%%%%%%%%%%%%%%%%%%%%%%%%%%%%

\no Interactions between the ${e}-${\it sector}
and the ${\bf \mu }$ one consisting of the two branches
of the $({e-\mu })$ fluid are introduced
via three mechanisms: (i) Through the choice of
the moduli parameters, (ii) the distribution of
the fractional $D0$-branes on the droplet and (iii) via
a gauge principle.

Concerning the first contribution to interactions,
the point is that because of the presence of the
${e}${\it --sector}, the particles of the
${\mu}${\it --sector} (the muons) will feel
not only the external magnetic field $B=B_{1}$
but also an induced term coming from the charged particles
of the ${e}${\it --sector}. As such electrons of
the ${\bf \mu }${\it --sector} view a total magnetic
field
\beq
B^{\ast }=B_{2}=B_{1}+\Delta B.
\eeq
From previous analysis, see~(\ref{equiv},\ref{n3ncon}),
one learns that $B_{2}=\frac{k_{2}}{k_{1}}B_{1}$.
The latter relation is based on the identity
$\theta _{1}=\theta _{2}\equiv \theta $.

For the distribution of the $D0$-branes of
the $\nu_{k_1k_2}$
FQH states, we suppose that the constraint equations
$J_{0}^{(i)}|\Phi\rangle =N_{i}\nu _{k_i}^{-1}|\Phi\rangle$
established for the Laughlin states are general ones and so
we demand that such a condition is also valid for FQH
states that are not of the Laughlin type. Put differently,
we will suppose that the total number of the fractional
$D0$-branes in the vacuum configuration $|\Phi \rangle $
of the $\nu_{k_{1}k_{2}}$ FQH states is
given by $J_{0}|\Phi \rangle =\frac{N}{\nu }|\Phi\rangle$; i.e
\begin{equation}
\lb{tnfd0b}
J_{0}|\Phi \rangle =\frac{k_{1}k_{2}}{k_{1}+k_{2}}N
|\Phi \rangle
\end{equation}
where the $J_{0}$ charge operator is
the full charge operator to be given
later.

The third contribution to interactions between
the two branches of the $(e-\mu)$ fluid comes
from the requirement that the ${e}$ and ${\bf \mu }$
couplings are $U(N_{1})\times U(N_{2})$ gauge invariant.
From the $U(N_{1})\times U(N_{2})$ group representation
analysis~(\ref{gstr}) and~(\ref{grep}), one
sees that the candidate fields to carry
such interactions behave as
\beq
\bra{l}
\lb{cafi}
\Psi _{\alpha a}\ \
\sim ({\bf N}_{1},{\bf N}_{2}) \\
{\bar{\Psi}}^{\bar{\alpha}\bar{a}}\ \
\sim ({\bf {\bar{N}}}_{1},{\bf {\bar{N}}}_{2}) \\
{\Psi }_{\alpha }^{\bar{a}}\ \
\sim ({\bf N}_{1},{\bf {\bar{N}}}_{2})\\
{\bar{\Psi}}^{\bar{\alpha}}_a\ \
\sim ({\bf {\bar{N}}}_{1},{\bf N}_{2}).
\era
\eeq
Therefore, there are two kinds of rectangular
complex matrices $\Psi_{\alpha a}$\ and
${\Psi }_{\alpha }^{\bar{a}}$\ together with their
complex conjugates. They look like the Polychronakos field,
but in fact they are more general objects with very remarkable
features. To get more insight in the role played by these fields,
let us focus our attention on one
of these fields, say the $\Psi _{\alpha a}$ and its conjugate
${\bar{\Psi}}^{\bar{\alpha} \bar{a}}$.
The results extend directly to the others.

%%%%%%%%%%%%%%%%%%%%%%%%%%%%%%%%%%%%%%%%%%%%%%%%%%%%%
\subsection{Interactions}
%%%%%%%%%%%%%%%%%%%%%%%%%%%%%%%%%%%%%%%%%%%%%%%%%%%%%
\no Now considering~(\ref{gstr}) and~(\ref{grep}),
and restricting to the $\Psi _{\alpha a}$
and ${\bar{\Psi}}^{\bar{\alpha}\bar{a}}$ fields,
a possible $U(N_{1})\times U(N_{2})$ gauge invariant
interacting action ${\cal S}_{int}$ one can write
down, up to the fourth order in the fields, is
\beq
\bra{l}
\lb{inact}
{\cal S}_{int} =\frac{i}{2}\int dt\;
\left( {\bar{\Psi}}^{\bar{\alpha}\bar{a}}
\partial _{t}\ \Psi _{\alpha a}+
{\bar{\Psi}}^{\bar{\alpha}\bar{a}}
{A_{1}}_{\alpha }^{\bar{\beta}} \Psi _{\beta a}+
 {\bar{\Psi}}^{\bar{\alpha}\bar{a}}
{A_{2}}_{a}^{\bar{b}} \Psi _{\alpha b}+hc\right)  \\
\qquad\;\;
+\frac{i}{2}\int dt\left( g_{1}
{\bar{\Psi}}^{\bar{\alpha}\bar{a}}
{Z_{1}}_{\alpha }^{\bar{\beta}}\Psi _{\beta a}+
g_{2} {\bar{\Psi}}^{\bar{\alpha}\bar{a}}Z_{2a}^{\bar{b}}
\Psi _{\alpha b}+hc\right) \\
\qquad\;\;
+\frac{i}{2}\int dt\;
\left( g_{3} {\bar{\Psi}}^{\bar{\alpha}\bar{a}}
{\Psi _{1}}_{\alpha }{\bar{\Psi}_{1}}^{\bar{\beta}}
\Psi _{\beta a}+g_{4}{\bar{\Psi}}^{\bar{\alpha}\bar{a}}
\Psi _{2a}{\bar{\Psi}_{2}}^{\bar{b}}\Psi _{\alpha b}+g_{5}
{\bar{\Psi}}^{\bar{\alpha}\bar{a}} {Z_{1}}_{\alpha }^{\bar{\beta}}
Z_{2a}^{\bar{b}}\Psi _{\beta b}+hc\right).
\era
\eeq
In this relation\ the $SU\left( N_{1}\right) $ indices $\alpha$,
$\bar{\alpha }$ $\left({\rm{resp.}}\; SU\left( N_{2}\right)
{\rm{indices}}\; a,\bar{a}\right)$ are contracted and
the summation over the range
$1\leq \alpha =\bar{\alpha }\leq N_{1}$
$\left({\rm{ resp}}.\; 1\leq a=\bar{a}\leq N_{2}\right) $
is understood. The five $g_{i}$
parameters are special coupling constants
involving a product of the $\Psi_{\alpha a}$ field
and its conjugate; there exist other coupling parameters
which are not important for the forthcoming
analysis and which we have set
to zero for simplicity. Note in passing that
due to the $\Psi _{\alpha a}$ field, the term
$g_{5}{\bar{\Psi}}^{\bar{\alpha}\bar{a}} {Z_{1}}_{\alpha}^{\bar{\beta}}
 Z_{2a}^{\bar{b}}\Psi _{\beta b}$ involves couplings of the
${Z_{1}}$ and ${Z_{2}}$ matrix variables already at the
fourth order in the fields, while one needs to go to the sixth
power if one is using only the Polychronakos type fields as shown
in the following interacting term, ${\bar{\Psi}}_{1}^{\bar{\alpha}}
{\bar{\Psi}}_{2}^{\bar{a}}\ {Z_{1}}_{\alpha }^{\bar{\beta}}
Z_{2a}^{\bar{b}}\Psi _{1\beta }\Psi _{2b}$.

The constraint equations one gets from the full
action ${\cal S}$ of the $({e}-{\mu})$
system contain extra contributions coming from
the interacting part~(\ref{inact}). The new quantum
constraint equations read therefore as
\beq
\bra{l}
\lb{nq4ncon1}
J_{\alpha \bar{\alpha}}^{(1)}|\Phi \rangle =0\\
J_{0}^{(1)}|\Phi\rangle =k_{1}N_{1}|\Phi\rangle\\
J_{a\bar{a}}^{(2)}|\Phi \rangle =0\\
J_{0}^{(2)}|\Phi \rangle=k_{2}N_{2}|\Phi\rangle
\era
\eeq
where now the $U(N_{1})\times U(N_{2})$
currents have a $\Psi _{\alpha a}$
and ${\bar{\Psi}}_{\bar{\alpha}{\bar a}}$
dependence as shown below
\beq
\bra{l}
\lb{nq4ncon2}
J_{\alpha \bar{\alpha}}^{(1)} =
[Z_{1},{\bar{Z}_{1}}]_{\alpha \bar{\alpha}}+
\frac{\theta }{2k_{1}}\left(\Psi _{1\alpha }
{\bar{\Psi}}_{1{\bar{\alpha}}}
+ \sum_{a=1}^{N_{2}}
\Psi _{\alpha a}{\bar{\Psi}}_{{\bar\alpha} {\bar a}}-
J_{0}^{(1)}\delta _{\alpha \bar{\alpha}}\right) \\
J_{0}^{(1)}  = \sum_{\alpha =1}^{N_{1}}
\left( {\bar{\Psi}}_{1\bar{\alpha}} \Psi _{1\alpha }+
\sum_{a=1}^{N_{2}}
{\bar{\Psi}}_{{\bar\alpha} {\bar a}} \Psi _{\alpha a}\right)
\era
\eeq
and similar relations for the $\mu${\it --sector}
\beq
\bra{l}
\lb{nq4ncon3}
J_{a\bar{a}}^{(2)} =[Z_{2},{\bar{Z}_{2}}]_{a\bar{a}}+
\frac{\theta }{2k_{2}}\left(\Psi _{2a} {\bar{\Psi}}_{2\bar{a}}
+\sum_{\alpha =1}^{N_{1}}
\Psi_{\alpha a}{\bar{\Psi}}_{\bar{\alpha} {\bar a}}
-J_{0}^{(2)}\delta _{a\bar{a}}\right) \\
J_{0}^{(2)} =\sum_{a=1}^{N_{2}}
\left( {\bar{\Psi}}_{2\bar{a}} \Psi _{2a}+
\sum_{\alpha =1}^{N_{1}}{\bar{\Psi}}_{\bar{\alpha} {\bar a}}
\Psi _{\alpha a}\right).
\era
\eeq
It is interesting to note here that as far
as consistency of the matrix
model is concerned, one does not need to
introduce all the different kinds
of the $\Psi $'s we have considered above.
It is possible to accomplish our objective
with the $\Psi _{\alpha a}$ field in the
bi-fundamental of the
$U(N_{1})\times U(N_{2})$ gauge group.

%%%%%%%%%%%%%%%%%%%%%%%%%%%%%%%%%%%%%%%%%%%%%%%%%%%%%
\subsection{$(e-\mu)$--Matrix model}
%%%%%%%%%%%%%%%%%%%%%%%%%%%%%%%%%%%%%%%%%%%%%%%%%%%%%

\no In this special matrix model, the
$\Psi _{1\alpha }$ and $\Psi _{2a}$\
Polychronakos fields and the
${\Psi }_{\alpha }^{\bar{a}}$ field are ignored
and \ so the effective variables of the
$(e-\mu)$ system are reduced to
$\left\{ Z_{1},Z_{2},A_{1},A_{2},\Psi _{\alpha a}\right\} $,
while the total action ${\cal S}$, obtained
from~(\ref{inact}) by setting also
$g_{1}=g_{2}=0$, reads as
\beq
\bra{l}
\lb{ninact}
{\cal S} =\int dt\;
\sum_{i=1}^{2}\Big[ {\frac{k_{i}}{4\theta }}
{\rm{Tr}}\left( i{ \bar{Z}_{i}}DZ_{i}-
\omega {\bar{Z}_{i}}Z_{i}+2\theta A_{i}\right)
\Big] +hc \\
\;\;\; +\int dt\;
\left[\frac{i}{2}{\bar{\Psi}}^{\bar{\alpha}\bar{a}}
\left(\partial _{t}+{A_{1}}_{\alpha }^{\bar{\beta}}
\delta _{a}^{\bar{b }}+{A_{2}}_{a}^{\bar{b}}
\delta _{\alpha }^{\bar{\beta }}\right)
\Psi _{\bar{\beta} \bar{b}}+
\lambda {\bar{\Psi}}^{\bar{\alpha}\bar{a}}
{Z_{1} }_{\alpha }^{\bar{\beta}} Z_{2a}^{\bar{b}}
\Psi _{\beta b}\right]+ hc
\era
\eeq
where we have set $\omega _{1}=\omega _{2}=\omega $.
The above realization of the $J_{\alpha \bar{\alpha}}^{(1)}$
and $J_{a\bar{a}}^{(2)}$ currents~(\ref{nq4ncon2},\ref{nq4ncon3})
simplifies to
\beq
\bra{l}
\lb{nq4ncon4}
J_{\alpha \bar{\alpha}}^{(1)}  =
 [Z_{1},{\bar{Z}_{1}}]_{\alpha \bar{\alpha}}+
 \frac{\theta }{2k_{1}}\left( \sum_{a=1}^{N_{2}}
\Psi _{\alpha a}{\bar{\Psi}}_{ \bar{\alpha} {\bar a}}
-J_{0}^{(1)} \delta _{\alpha \bar{\alpha}}\right) \\
J_{a\bar{a}}^{(2)}  =
[Z_{2},{\bar{Z}_{2}}]_{a\bar{a}}+\frac{\theta }
{2k_{2}} \left( \sum_{\alpha =1}^{N_{1}}
\Psi _{\alpha a}{\bar{\Psi}}_{\bar{\alpha}{\bar a} }-
J_{0}^{(2)}\delta _{a\bar{a}}\right)
\era
\eeq
while the two $U(1)$ charge operators $J_{0}^{(1)}$ and
$J_{0}^{(2)}$ reduce to
\beq
\bra{l}
\lb{chop}
J_{0}^{(1)}  = \sum_{\alpha =1}^{N_{1}}
\left( \sum_{a=1}^{N_{2}}\ \
{\bar{\Psi}}_{\bar{\alpha}{\bar a}} \Psi_{\alpha a} \right) \\
J_{0}^{(2)}  = \sum_{a=1}^{N_{2}}
\left( \sum_{\alpha =1}^{N_{1}}\ \
{\bar{\Psi}}_{\bar{\alpha} {\bar a}} \Psi_{\alpha a}\right) .
\era
\eeq
Comparing these two relations, one discovers,
upon interverting the sums $\sum_{\alpha =1}^{N_{1}}$ and
$\sum_{a=1}^{N_{2}}$, that  the two $U(1)$
charge currents are equal, i.e $J_{0}^{(1)}=J_{0}^{(2)}$.
So the number $ k_{1}N_{1}$ of quasi-electrons and
the number $k_{2}N_{2}$\ of quasi-muons in the vacuum
configuration of the $(e-\mu)$ droplet should be equal;
that is we should have the equality
\beq
\lb{iden}
k_{1}N_{1}=k_{2}N_{2}.
\eeq
This constraint
equation is not a strange relation; it is in
fact expected from group theoretical
analysis of the vacuum wavefunction
and a result of subsection (2.2). Indeed,
due to non-commutative geometry, the
$SU\left( k_{1}\right) $ and $SU\left(k_{2}\right) $
singularities of the Laughlin wavefunctions of the ${e}$
and $\mu $ {\it sectors} are removed and therefore
the classical $\Psi $ field in the bi-fundamental~(\ref{cafi})
should be replaced, at the quantum level, by
the $\Psi_{\alpha i_{\alpha },aj_{a}}^{\pm }$
operators transforming in
the $\left( {\bf N}_{1}{\bf ,k}_{1}\right)
\otimes \left( {\bf N}_{2}{\bf ,k} _{2}\right) $
representation of the $\Big( SU\left( N_{1}\right) \times
SU\left( k_{1}\right) \Big) \otimes
\Big( SU\left( N_{2}\right) \times
SU\left( k_{2}\right) \Big) $ group.
Since the indices $\alpha i_{\alpha }$
and $aj_{a}$ are paired, invariance under
$\Big( SU\left( N_{1}\right)
\times SU\left( k_{1}\right) \Big) \otimes
\Big( SU\left( N_{2}\right)
\times SU\left( k_{2}\right) \Big) $
requires that we should have the
identity~(\ref{iden}).
Moreover using the constraint~(\ref{tnfd0b}), one
gets the following remarkable relation
\begin{equation}
\lb{rere}
\frac{N_{1}}{\nu _{1}}=\frac{N_{2}}{\nu _{2}}\ev
\frac{N_{1}+N_{2}}{\nu
_{1}+\nu _{2}}=\frac{N}{\nu }
\end{equation}
which is solved by
\beq
\bra{l}
N_{1}  = \frac{k_{2}}{k_{1}+k_{2}}N \\
N_{2}  = \frac{k_{1}}{k_{1}+k_{2}}N.
\era
\eeq
For the example of the $\nu =\frac{2}{5}$
FQH state, consistency requires
that the number $N_{1}$ of electrons
should be such that $N_{1}=\frac{5}{6}N$
and the number $N_{2}$ of muons is given
by $N_{2}=\frac{1}{6}N.$ Note
that the number $N_{2}$ coming from
the second condensation is smaller than
the number $N_{1}$ of electrons coming
from the first condensation. This
property is suspected to be valid for higher
orders of the hierarchy; i.e.,
$ N_{1}>N_{2}>N_{3}>...$

The wavefunction $|\Phi \rangle $ describing
the $(e-{\mu})$
system of $N$ electrons ($N_{1}$ electrons
and $N_{2}$ muons) on the NC
plane ${\RR}_{\theta }^{2}$ with
filling factor $\nu_{k_{1}k_{2}}$
should obey the constraint
equations~(\ref{tnfd0b},\ref{nq4ncon1})
and~(\ref{nq4ncon4},\ref{chop}).
Once we know the fundamental state
$|\Phi _{\nu_{k_{1}k_{2}}}^{(0)}\rangle $,
excitations are immediately determined by applying the
usual rules. Upon recalling the quantum coordinates as
\beq
\bra{l}
Z_{1\alpha {\bar{ \alpha}}}=
\sqrt{\frac{\theta }{2}}r_{\alpha {\bar{\alpha}}}^{+}
\\
Z_{2a{ \bar{a}}}=\sqrt{\frac{\theta }{2}}s_{a{\bar{a}}}^{+}
\era
\eeq
the total Hamiltonian ${\cal H}$ of the $({e}-{\mu })$
system, which contain two parts $ {\cal H}_{0}$ and
${\cal H}_{int}$, may be treated as the sum of a free part
given by
\beq
\lb{ham1}
 {\cal H}_{0}={\omega\ov 2}
\Big(2 {\cal N}_{e}+ 2{\cal N}_{\mu }+
N_{1}^{2}+N_{2}^{2} \Big)
\eeq
where
\beq
\bra{l}
{\cal N}_{e}=
\sum_{\alpha ,\beta=1}^{N_{1}}
r_{\alpha \beta }^{\dagger }r_{\beta \alpha }^{-}
\qquad \\
{\cal N} _{\mu }=
\sum_{a,b=1}^{N_{2}}s_{ab}^{\dagger }s_{ba}^{-}
\era
\eeq
are the operator numbers counting the ${e}$ and
${\mu }$ particles respectively, and
an interacting part
\beq
\lb{ham2}
{\cal H}_{int}\sim \left( {\psi }_{\bar{a}\bar{\alpha} }^{+}
r_{\alpha \bar{{\beta }}}^{+}\ s_{a\bar{{b}}}^{-}
\psi_{\beta b}^{-}+hc\right)
\eeq
describing couplings between the {\it two sectors}.
This interaction is a perturbation around ${\cal H}_{0}$
and so the spectrum of the full Hamiltonian may be obtained by
using standard techniques of
perturbation theory. The determination
of the vacuum configuration of $ {\cal H}_{0}$ depends,
however, on whether the NC geometry of the plane
is taken into account or not. In the simplest case
where the $SU\left(k_{1}\right) $ and
$SU\left( k_{2}\right) $\ symmetries of the singular
points are ignored, the creation and annihilation
operators $r_{\alpha {\bar{ \alpha}}}^{\pm }$,
$s_{a{\bar{a}}}^{\pm }$,\ and $\psi _{\alpha a}^{\pm }$
carry only the $SU\left( N_{1}\right)
\times SU\left( N_{2}\right) $ group
indices and so the Heisenberg algebra for these
operators reads as
\beq
\bra{l}
\lb{heisa}
\left[ \left( r^{-}\right) _{\alpha }^{\bar{\alpha }},
\left(r^{+}\right) _{\beta }^{\bar{\beta }}\right]
= \delta _{\alpha \beta}\delta ^{\bar{\alpha }
\bar{\beta }}  \\
\left[ \left( s^{-}\right) _{a}^{\bar{a}},
\left( s^{+}\right) _{b}^{ \bar{b}}\right]
= \delta _{ab}\delta ^{\bar{a}\bar{a}} \\
\left[ \left( \psi ^{-}\right) ^{\bar{\alpha }
\bar{a}},\left( \psi^{+}\right) _{\alpha a}\right]
= \delta _{\alpha }^{\bar{\alpha } }
\delta _{a}^{\bar{a}}\; 
\era
\eeq
while all others are given by commuting relations.
A way to build the spectrum of the Hamiltonian
${\cal H}_{0}$ with the constraint~(\ref{nq4ncon1})
is given with the aid of the special condensate operators
\beq
\left( A^{+}\right)_{a\alpha }^{\left(n,m\right) }=
\left( \left( s^{+}\right) ^{n-1}\psi^{+}
\left( r^{+}\right)^{m-1}\right) _{a\alpha }.
\eeq
The wavefunction for the vacuum reads , in terms
of the $\left( A^{+}\right)_{a\alpha }^{\left( n,m\right) }$'s,
as
\begin{equation}
\lb{wavef}
\left[ \varepsilon ^{\alpha _{1}...\alpha _{N_{1}}}
\prod_{j=1}^{p}O_{\alpha_{\left( jN_{2}+1\right) }...
\alpha _{\left( j+1\right) N_{2}}}^{\left(j\right) }
\right] ^{k_{1}}|0>
\end{equation}
where the $O^{\left( j\right) }$'s are building
blocks invariant under $SU\left( N_{2}\right) $ but
transforming as ${\bf N}_{1}^{\otimes N_{2}}$
under $SU\left( N_{2}\right) $ symmetry,
\begin{equation}
\lb{builb}
O_{\alpha _{\left( jN_{2}+1\right) }...
\alpha _{\left( j+1\right)
N_{2}}}^{\left( j\right) }=
\varepsilon ^{a_{\left( jN_{2}+1\right)}...
a_{\left( j+1\right) N_{2}}}\left( A^{+}\right) _{a_{\left(
jN_{2}+1\right) }
\alpha _{\left( jN_{2}+1\right) }}^{\left( 1,j\right)}...
\left( A^{+}\right) _{a_{\left( j+1\right) N_{2}}
\alpha _{\left(
j+1\right) N_{2}}}^{\left( N_{2},j\right) }.
\end{equation}
Invariance under $SU\left( N_{1}\right)
\times SU\left( N_{2}\right) $ is
ensured by considering $p$ building
blocks and applying the $SU\left(
N_{1}\right) $ antisymmetriser.
In addition to the manifest $SU\left(N_{1}\right)
\times SU\left( N_{2}\right) $ invariance, this configuration
has clearly
\beq
k_{1}N_{1}=pk_{1}N_{2}=k_{2}N_{2}
\eeq
charges $U\left( 1\right)$ and an energy
\beq
\lb{nener}
 E_{0}=k_{1}\left[ p\frac{\left( N_{2}-1\right) \left(
N_{2}-2\right) }{2}+\frac{\left( p-1\right)
\left( p-2\right) }{2}N_{2} \right] +
\frac{N_{1}+N_{2}}{2}.
\eeq
Effects of NC geometry of the plane may
be taken into account by considering
the splitting of the known
$SU\left(k_{1}\right)$ ({\rm{resp}}.
$SU\left( k_{2}\right)$)
singularities of the
Laughlin wavefunctions with filling factor
$\nu _{k_1}=\frac{1}{k_{1}}$
(\rm{resp}. $\nu _{k_2}=\frac{1}{k_{2}}$).
The previous $r_{\alpha {\bar{\alpha}} }^{\pm }$
and $s_{a{\bar{a}}}^{\pm }$ operators are now given by the sum
over $\left( {\bf r}^{\pm }\right) _{\alpha
i_{\alpha }}^{\bar{\alpha } i_{\bar{\alpha }}}$
and $\left( {\bf s}^{\pm }\right) _{ai_{a}}^{%
\bar{a}i_{\bar{a}}}$ as we have already
explained in section 2. The general canonical commutation
relations for the ($e-\mu$) matrix model, read now as
\beq
\bra{l}
\lb{gccr}
\left[ \left( {\bf r}^{-}\right) _{\alpha
i_{\alpha }}^{\bar{\alpha }i_{ \bar{\alpha }}},
\left( {\bf r}^{+}\right) _{\beta j_{\beta }}^{ \bar{\beta }
j_{\bar{\beta }}}\right]  = \delta _{\alpha \beta}
\delta ^{\bar{\alpha }\bar{\beta }}\delta _{i_{\al}j_{\beta}}
\delta ^{{ i}_{\bar\al}{j}_{\bar\beta}}  \\
\left[ \left( {\bf s}^{-}\right) _{ai_{a}}^{\bar{a}
i_{\bar{a} }},\left( {\bf s}^{+}\right) _{bj_{b}}^{\bar{b}
j_{\bar{b}}}\right]
 = \delta _{ab}\delta ^{\bar{a}\bar{a}}
\delta _{i_{\al}j_{\beta}}
\delta ^{{i}_{\bar\al}{j}_{\bar\beta}} \\
\left[ \left( {\bf \psi }^{-}\right) ^{\bar{\alpha }
i_{\bar{\alpha}}\bar{a}j_{\bar{a}}},
{\bf \psi }_{\alpha i_{\alpha }aj_{a}}^{+} \right]
= \delta _{\alpha }^{\bar{\alpha }}\delta _{a}^{\bar{a} }
\delta _{i_{\al}}^{{i}_{\bar\al}}
\delta _{j_{\beta}}^{{j}_{\bar\beta}}
\era
\eeq
and all the remaining others are identically zero.
In this case, one needs a
building block structure using invariants of
$SU\left( k_{1}\right) \times
SU\left( k_{2}\right) \times SU\left( N_{1}\right)
\times SU\left(N_{2}\right) $ symmetry.
The building blocks \ are constructed by using the
following: (i) $SU\left( k_{2}\right) $ invariants
involve $p$ factors of \ $ SU\left( k_{1}\right) $
invariants and (ii) $SU\left( N_{1}\right) $ scalars
need $p$ factors of $SU\left( N_{2}\right) $ condensate.
This property is based on the relations $k_{2}=pk_{1}$
and  $N_{1}=pN_{2}$.

%%%%%%%%%%%%%%%%%%%%%%%%%%%%%%%%%%%%%%%%%%%%%%%%%%%%%
\section{Conclusion}
%%%%%%%%%%%%%%%%%%%%%%%%%%%%%%%%%%%%%%%%%%%%%%%%%%%%%

In this paper we have developed a matrix
model for FQH states at filling factor
$\nu_{k_1k_2}$ going beyond
the Laughlin theory. To illustrate our
idea, we have considered a FQH system
of a finite number $N=\left(N_{1}+N_{2}\right) $
electrons with filling factor
$\nu_{k_{1}k_{2}}\ev
\nu_{p_{1}p_{2}}=\frac{p_{2}}{p_{1}p_{2}-1}$; $\ p_{1}$
is an odd integer and $p_{2}$ is an even integer.
The $\nu_{p_{1}p_{2}}$ series
corresponds just to the level two of
the Haldane hierarchy; it recovers the
Laughlin series $\nu_{p_{1}} =\frac{1}{p_{1}}$
by going to the limit $\ p_{2}$ large and contains several
observable FQH states of the
series $\frac{m}{2mp\pm 1}$, such as
those states with filling factor
$\nu = {2\ov 3}, {2\ov 5}, \cdots$
Our matrix model, which extends the
regularized Susskind theory considered by
Polychronakos for studying FQH
droplets, has a $U\left( N_{1}\right) \times
U\left( N_{2}\right) $ gauge
invariance and assumes that the
FQH fluid consists of two coupled branches
with filling fractors
$\nu_{k_{1}} =\frac{1}{k_{1}}$ and
$\nu_{k_{2}} =\frac{1}{k_{2}}$
where the $k_{1}$ and $k_{2}$ integers are
related to Haldane ones as $k_{1}=p_{1}$ and
$ k_{2}=k_{1}(p_{1}p_{2}-1)\equiv pk_{1}$.
The branch with
$\nu_{k_{1}}$
is the fundamental one and that with
$\nu_{k_{2}}$
is built on it. Couplings are
manifested through three different channels:
(1) Through the effective
external magnetic field
$B^{\ast }=\frac{k_{2}}{k_{1}}B$   felt by the
$ N_{2} $ electrons of the branch with
$\nu_{k_{1}}$;
here $B$ is the external magnetic field seen
by the $N_{1}$ electrons of
the fundamental branch with
$\nu_{k_{2}}$ and
responsible for the NC geometry of the plane.
(2)~by using a
natural hypothesis according to which
the total $U\left( 1\right) $ charge
of the LLL is equal to the product
of the inverse filling factor $\nu $
with the number $N$ of electrons. In other
words, the LLL fundamental wavefunction
$|\Phi >$ is subject to the
constraint equation
$J_{0}|\Phi >=\frac{N}{\nu }|\Phi >$,
where $J_{0}$\ is the $U\left( 1\right) $ charge operator.
Recall in passing that such a feature is required
as well for the case of the
Laughlin states with a finite number
of electrons. (3)~through the
involvement of a new field
$\Psi_{\alpha a}$ transforming in the
bi-fundamental representation
$\left( {\bf N}_{1}{\bf ,N}_{2}\right) $ of
the $U\left( N_{1}\right) \times U\left( N_{2}\right) $
gauge group. Such a
field connects the two branches of
the fluid, a feature illustrated by naively
consideing the field coupling
${\bar{\Psi}}^{\bar{\alpha}\bar{a}}
{Z_{1}}_{\alpha }^{\bar{\beta}}
Z_{2a}^{\bar{b}}\Psi _{\beta b}$
where ${Z_{1}} _{\alpha }^{\bar{\beta}}  $
and $Z_{2a}^{\bar{b}}$ are the
matrix field variables for the two branches,
respectively. The field $\Psi_{\alpha a}$ can
also accomplish the complete regularisation for a consistent
matrix model with a finite number of particles, without the
need to introduce the Polychronakos fields $\Psi _{\alpha }$
and $\Psi _{a}$.   In this special
case the $U\left( 1\right) $ charge
operator reduces to\\
$$J_{0}=\sum_{\alpha=1}^{N_{1}}
\sum_{a=1}^{N_{2}}\ \
{\bar{\Psi}}_{\bar{\alpha} \bar{a} } \Psi _{\alpha a}$$.\\
The $\Psi _{\alpha a}$ field may be also viewed
as the first element of
a series of fields in the fundamentals
$\left( {\bf N}_{1},...,{\bf N} _{n}\right) $
of the $\left( \otimes _{i=1}^{n}U\left( N_{i}\right) \right) $
gauge group of a multi-component fluid
matrix model. Recall that the case we
have studied here is in fact just a particular
FQH states of a more general
one where the fluid droplet is assumed
to consist of several coupled
branches, say $n$ branches, with a
filling factor\\
$$
\nu =\frac{\sum_{i=1}^{n}k_{i}}
{\prod_{j=1}^{n}k_{j}}.
$$ \\
$n=1$ is the Laughlin model, $ n=2$ is the model
we have discussed here and $n\geq 3$ is the generic case.

In this paper we have also studied an
interesting feature of singularities in
spaces with a NC geometry. This special
property has not been addressed
before in the context of FQH systems.
The point is that the Laughlin
wavefunctions~(\ref{lau})
with filling factor
$\nu =\frac{1}{k}$  have
degenerate zeros as sources of singularities
of type $A_{r}$. A simple way to
see it is to go to the limit
\\
$$z_{\beta }=z_{\alpha }+\epsilon _{\alpha\beta}$$\\
and substitute in the above $\Phi _{L}$~(\ref{lau}). One
obtains a product of monomials $%
\epsilon _{\alpha\beta }^{k}$ which behave as an
$SU\left( k\right) $ singularity on
the plane. However, due to the presence
of the external magnetic field $B$, the
two-space is no longer commutative and
one expects this kind of
singularity to be removed in agreement
with the general property of the absence
of singularities in NC spaces.
On the basis of established results in a NC
geometry context, especially varieties with
discrete symmetries as it is the
case for the Laughlin wavefunctions, we have
completed previous partial
results obtained recently by taking
into account the effect of NC geometry.
As a consequence, the spectrum involves
now a larger symmetry group, namely
the $U\left( N_{1}\right) \times U\left( N_{2}\right) $
gauge group of the
matrix model, but also the
$SU\left( k_{1}\right) \times SU\left(
k_{2}\right) $ living at the singular
points. One of the striking results we
have obtained in this issue is that real
electrons are $D0$-branes behaving
as singlets of the $SU\left( k_{1}\right)
\times SU\left( k_{2}\right) $
group. They are composite objects of
elementary excitations transforming in
fundamental representations of the
above groups and behave exactly as
the known fractional $D0$-branes at
singularities one has in brane physics.
We have presented the essentials about these
fractional $D0$-branes versus FQH fluid
droplets, but further insight is needed
to obtain the general picture.

To end this conclusion, we would like to note that general
solutions involving several Pohychronakos fields have been derived
in~\cite{morariu}. There it was shown that starting from a matrix
model of a FQH state with filling factor $\nu={1\ov k}$, and
replacing
the usual Polychronakos action term $S_{\Psi}$ by one expressed in
terms of a
Wilson line gauge field, one gets interesting results with several
applications, in particular for the study of multilayer
quantum Hall fluids. This important development seems to have a close
link with the approach we have developed in this paper, especially
the part concerning the study of FQH states of rational filling
factor that are not of Laughlin type. Despite this link, we think
that the two methods differ in other aspects, in particular when
one deals with layers with different filling factors. In
addition to the fact that the Wilson line generalization
of~\cite{morariu} involves only one integer $k$, contrary to our
study, an extension of the analysis of~\cite{morariu} may also be
worked out as a generalization of the model we have developed in this
paper. Details of this direction will be reported elsewhere.

\section*{Acknowledgements}%--------------------

A. Jellal and E.H. Saidi would like to thank Professor L. Susskind
for several discussions. AJ~is grateful to Professors P.
Bouwknegt, \"O.F. Dayi, G. Ortiz, T. Palev and R.A.~R\"omer for
their encouragments. EHS~is thankful to the organizers of the
Workshop on "Quantum Gravity, String Theory and Cosmology"
(February 4-22, 2002) held at Stellenbosch University, South
Africa. He is also thankful to the Stellenbosch Institute for
Advanced Study (STIAS) for kind hospitality.

\end{document}